\documentclass[draft]{agujournal2019}
\usepackage{url} %this package should fix any errors with URLs in refs.
\usepackage{lineno}
\usepackage[inline]{trackchanges} %for better track changes. finalnew option will compile document with changes incorporated.
\usepackage{soul,color}
\usepackage{graphicx}
\usepackage{subcaption}
\draftfalse
\journalname{Geophysical Research Letters}

\begin{document}

\title{Deep Learning-based Epicenter Localization\\using Single-Station Strong Motion Records}

%  AUTHORS AND AFFILIATIONS
\authors{Melek~Türkmen\affil{1},
        Sanem~Meral\affil{2},
        Baris~Yilmaz\affil{1},
        Melis~Cikis\affil{3},
        \\
        Erdem~Akagündüz\affil{1}, 
        Salih~Tileylioglu\affil{4}}
\affiliation{1}{Dept. of Modeling and Simulation, Graduate School of Informatics, METU, Türkiye}
\affiliation{2}{Dept. of Systems Engineering, Turkish Aerospace Inc., Türkiye}
\affiliation{3}{Dept. of Systems Engineering, ASELSAN Inc., Türkiye}
\affiliation{4}{Dept. Civil Engineering, Kadir Has University, Türkiye}

% Corresponding author mailing address and e-mail address:
\correspondingauthor{Melek Türkmen}{turkmen.melek@metu.edu.tr}

% KEY POINTS
\begin{keypoints}
\item In this paper, deep learning techniques are applied to strong motion records in order to locate epicenters at single stations.
\item This study examines whether strong motion records, which are rarely used for seismology-related studies, contain information about an earthquake's characteristics, and whether DL-based methods can benefit from them.
\item We introduce a large-scale strong motion record collection, AFAD-1218, which contains over 36,000 strong motion records from Turkey.
\end{keypoints}

%% \begin{abstract} starts the second page
\begin{abstract}
{This paper explores the application of deep learning (DL) techniques to strong motion records for single-station epicenter localization. Often underutilized in seismology-related studies, strong motion records offer a potential wealth of information about seismic events. We investigate whether DL-based methods can effectively leverage this data for accurate epicenter localization. Our study introduces AFAD-1218, a collection comprising more than 36,000 strong motion records sourced from Turkey. To utilize the strong motion records represented in either the time or the frequency domain, we propose two neural network architectures: deep residual network and temporal convolutional networks. Through extensive experimentation, we demonstrate the efficacy of DL approaches in extracting meaningful insights from these records, showcasing their potential for enhancing seismic event analysis and localization accuracy. Notably, our findings highlight significant reductions in prediction error achieved through the exclusion of low signal-to-noise ratio records, both in nationwide experiments and regional transfer-learning scenarios. Overall, this research underscores the promise of DL techniques in harnessing strong motion records for improved seismic event characterization and localization.}
\end{abstract}

%  BODY TEXT

\section{Introduction}
\label{sec:introduction}
The sudden rupture of faults beneath the earth’s surface cause seismic waves that propagate towards the earth’s surface. The waves reaching the ground surface may cause weak to strong shaking of the ground called earthquakes (EQ). Vibrations resulting from earthquakes are recorded by instruments placed at various locations on the Earth’s surface. These recordings are used by seismologists and engineers to better understand earthquakes and their effects on the built environment. 

{The expanding number of available earthquake recordings has lead to seismic source  parameters such as origin time, location, and magnitudes of individual earthquakes being calculated automatically using various computational algorithms and thereby increasing the speed and accuracy of the estimated parameters}. Furthermore, the rise in the number of recordings has also paved the way for applying data-driven methods to estimate source parameters. The availability of various artificial intelligence methods and their successful application to different areas in science coupled with the volume of earthquake recordings has allowed the number of research in this area to grow exponentially in recent years. Moreover, there is also curiosity on earthquake features that may be extracted from data using advanced AI methods. Most of the recent research covers a variety of topics including event detection, epicenter localization, magnitude detection, earthquake early warning systems (EEW) \cite{Mousavi2023}.

%\hl{P2: Recent literature... This paragraph ML based...} \\ \\ \\ \\ \\ \\ \\ \\ 
The earlier AI methods used in seismology and earthquake engineering included various machine learning techniques (ML) such as artificial neural networks (ANN), support vector machines (SVM), and decision trees \cite{Pengcheng2020}. \cite{Böse2008}, \cite{Bose2012} proposed multi-station and single-station-based ANN algorithms respectively, to determine earthquake sources and ground motion parameters to be used as a part of EEW systems. \citeA{Ochoa2018} built a Support Vector Machine Regression (SVMR) model to implement an EEW for the city of Bogota using recordings from a single station. Recently, \citeA{YANG2021} used conventional machine learning techniques as well as Convolutional Neural Networks (CNN) to distinguish between deep and shallow micro-seismic events. In another recent study, \citeA{McBrearty2022} uses graphical neural networks (GNN) to estimate the location and magnitude of earthquakes and compare to traditional methods. 

ML algorithms used in the studies above are confined to modeling and generalizing limited-scale data sets due to their reliance on fixed, hand-crafted feature representations. The advantage of deep learning algorithms, on the other hand, is their ability to automatically learn hierarchical representations directly from earthquake recordings, which enables them to process high-dimensional and complex datasets more efficiently. 

%this is the DL paragraph
The DL-based approaches to earthquake signal processing are pioneered by \citeA{Perol2018}, which proposes a CNN-based model for rough localization (i.e. classifying into region clusters) and event detection, based on a relatively small-scale dataset ($\sim$3k events). Following this study, DL-based approaches have increasingly been utilized in 
event detection for EEW \cite{Kuyuk2018,Lomax2019,Mousavi2020b,Münchmeyer2020,Yano2021,Bilal2021},
event classification \cite{Kim2022,Nakano2022}, 
ground response estimation \cite{Hong2021},
EQ phase picking \cite{Pardo2019,Mousavi2020b},
magnitude estimation \cite{Jozinović2020,Ende2020,Münchmeyer2020,Zhang2021,Ristea2022,Saad2021,Bloemheuvel2022}, 
EQ origin time estimation \cite{Mousavi2020a,Saad2021}, 
epicenter location classification \cite{Kuyuk2018,Lomax2019,Kriegerowski2019,Saad2021}, 
epicentral distance estimation \cite{Ristea2022,Yoma2022},
depth prediction \cite{Kriegerowski2019,Mousavi2020a,Ende2020,Zhang2021,Ristea2022,Saad2021,Bilal2021}, 
and epicenter coordinates prediction \cite{Zhang2020,Mousavi2020a,Ende2020,Zhang2021,Bilal2021}.  

All the above-mentioned studies, while addressing different problems of earthquake engineering, have in common the fact that they aim to learn hierarchical representations directly from a relatively large corpus of seismic waveforms. Several of these studies \cite{Kuyuk2018, Lomax2019, Zhang2020, Zhang2021, Saad2021, Sugiyama2021, Yoma2022, Nakano2022} utilize relatively medium-scale local datasets ($<$5k earthquake events) as inputs to their experiments, similarly to \cite{Perol2018}.
The question of whether such a scale of data is sufficient for building high-level representations in DL models is difficult to answer.  A large-scale corpus containing hundreds of thousands of EQ events did not exist until very recently. To meet this demand, \cite{Mousavi2019}  published the Stanford Earthquake Dataset (STEAD), which included more than a million three-component waveforms that belong to  $\sim$450k recorded seismic events. The same group used a subset of this dataset in their subsequent studies, first using a Bayesian DNN to detect P-arrival times and localize epicenters \cite{Mousavi2020a}, and then using a transformer and LSTM-included architecture for event and phase picking \cite{Mousavi2020b}. %In addition, another group \cite{Ristea2022} interested in this large-scale dataset used a complex CNN to solve the earthquake localization and magnitude estimation problems, and reported a slight improvement. 
Some other studies also utilized large-scale sets, such as \cite{Pardo2019,Nakano2022,Kim2022}, but not for earthquake localization as we intend in this paper, but rather for phase picking or event classification. 

In seismology and engineering, two types of instruments play a significant role in providing the necessary dataset scale for DL-based applications: seismometers and accelerometers. Seismometers measure ground velocity, offering insights into seismic wave characteristics and earthquake magnitude. These sensors constitute the majority of large-scale datasets found in the literature, including the STEAD dataset, primarily because seismometers are capable of detecting long-distance vibrations such as earthquakes or volcano tremors. Accelerometers, on the other hand, measure ground acceleration, providing detailed information on the intensity of shaking experienced by structures. While seismometer records contribute to understanding earthquake propagation, accelerometer records are essential for evaluating structural response and guiding earthquake-resistant design practices. Despite not being typically utilized for seismology-focused studies in the literature, questions arise regarding whether these strong motion records contain valuable information about an earthquake's characteristics and whether DL-based methods can learn from them.

\begin{figure}[h]
\centering
    \includegraphics[trim=37 13 43 20,clip=true,width=0.8\textwidth]{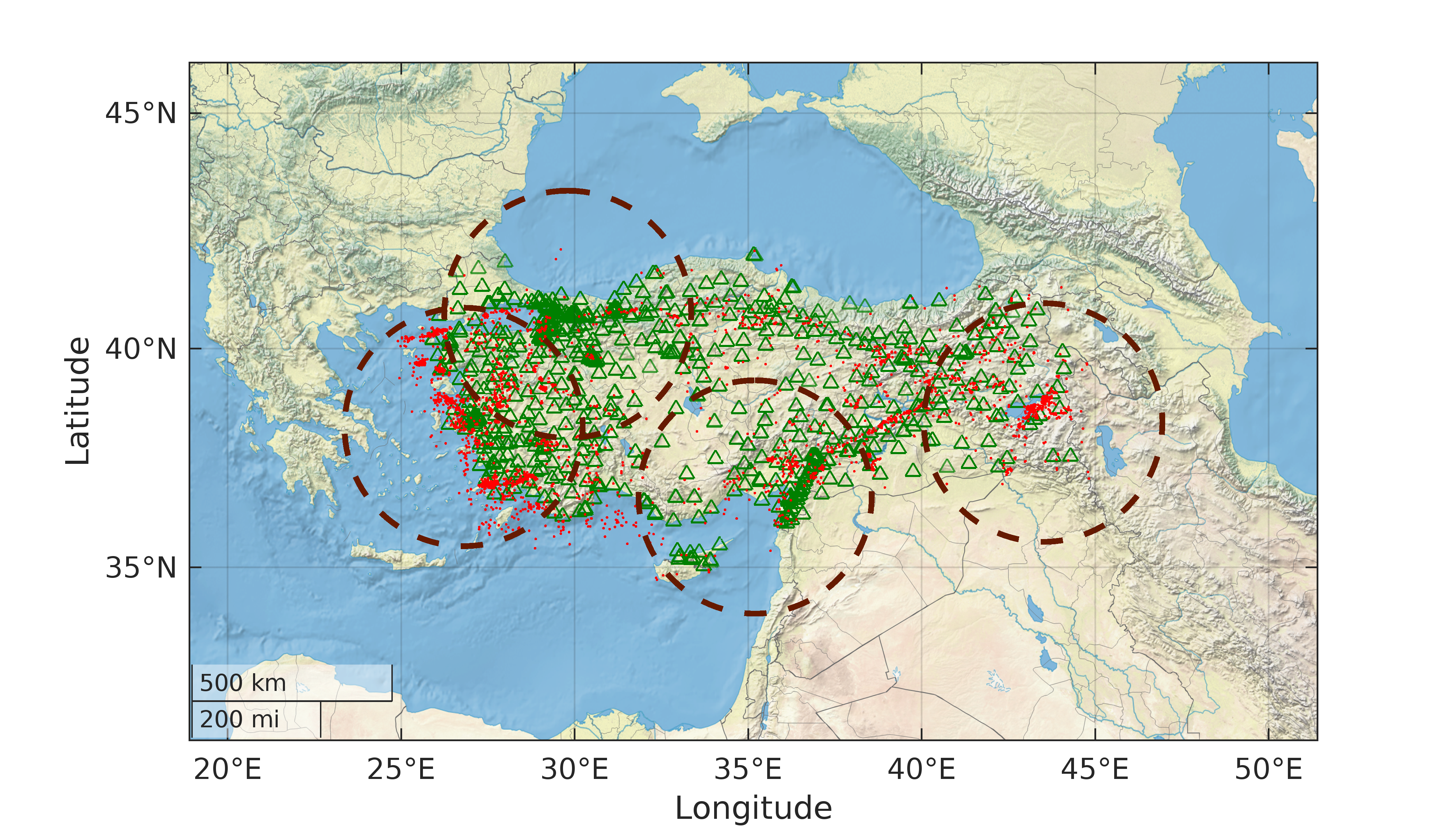}
\caption{A total number 3,655 events (red dots), collected from 718 stations (green triangles), exist in the AFAD-1218 dataset. Transfer learning experiments are conducted within the four regions (West, NorthWest, South, East) delineated by dashed circles.}
\label{fig:AFAD-EQs}
\end{figure}

\subsection{Problem Definition}
From a machine learning standpoint, the aforementioned DL-based approaches all utilize supervised algorithms and can be broken down into three main problem categories, namely detection, classification, and regression. Detection-based studies aim to identify the earthquake event \cite{Kuyuk2018,Lomax2019,Mousavi2020b,Münchmeyer2020,Yano2021,Bilal2021}, or the phase instant \cite{Pardo2019,Mousavi2020b}, such as the primary (P) or secondary (S) wave starting time. Classification-based studies, on the other hand, are either an extension of a detection problem, such as classifying an event into multiple categories such as earthquake, tremor, or noise \cite{Kim2022,Nakano2022}; or a simplification of a regression problem such as localizing an epicenter into pre-clustered regions \cite{Kuyuk2018,Lomax2019,Kriegerowski2019,Saad2021}.  Regression-based approaches aim at estimating a parameter of an event such as the origin time \cite{Mousavi2020a,Saad2021}, epicentral distance \cite{Ristea2022,Yoma2022}, epicentral orientation \cite{Lomax2019,Saad2021}, epicentral coordinates \cite{Zhang2020,Mousavi2020a,Ende2020,Zhang2021,Bilal2021}, depth \cite{Kriegerowski2019,Mousavi2020a,Ende2020,Zhang2021,Ristea2022,Saad2021,Bilal2021} or magnitude \cite{Jozinović2020,Ende2020,Münchmeyer2020,Zhang2021,Ristea2022,Saad2021,Bloemheuvel2022}. Undoubtedly, regression problems are more challenging to solve, and hence they cover the most recent literature. In this paper, we specifically focus on the epicentral coordinates prediction problem, basically finding the projection of the focus of the earthquake on the surface of the earth in world coordinates. Coordinate prediction differs from epicentral distance estimation, which is finding the distance from the receiving sensor to the epicenter, or epicentral orientation prediction, which is finding the angular orientation of the epicenter relative to the receiving sensor location; however, it is a superset of both. Hence, it is considered a complete problem definition for epicenter localization. 

There are a considerable number of studies that attack the epicentral location in world coordinates problem in the literature. The majority of these studies \cite{Kriegerowski2019,Zhang2020,Ende2020,Zhang2021,Bilal2021} utilize a limited-sized dataset or rely on auxiliary information such as seismic phase arrival times and/or multiple stations distribution within a network. Although there is one study \cite{Mousavi2020a} that utilizes a large-scale dataset suitable for a DL-based study, they use auxiliary information of P phase arrival times and study seismometer (broadband) sensors. In their recent work, \cite{Caglar2024} examine whether these models can effectively learn solely from raw seismometer records, without relying on auxiliary information such as seismic phase arrival times and station distribution within a network. 

In this paper, we focus on analyzing a large-scale dataset consisting solely of strong motion records and employing them with two distinct DL models for predicting epicentral coordinates. To the best of our knowledge, no previous studies have investigated the DL capabilities of strong motion records for earthquake-related tasks, particularly epicenter localization. Our study aims to determine the extent to which strong motion records can provide information about an earthquake's characteristics through DL techniques.
  
The representation domain of the raw time-series strong motion waveforms is a crucial consideration when feeding these signals into DL-based models. While the ML-based approaches heavily pre-process waveforms, DL-based approaches aim at building end-to-end models that rely on the raw waveform for their primary objective of developing hierarchical representations. Alternatively, the short-term Fourier transform (STFT), or spectrogram, of the signal is used in some DL-based studies  \cite{Ristea2022,Yoma2022,Nakano2022}  as the input, whereas the majority use the raw time signal. Regarding the behavior of a DL-model with respect to the domain of the input signal used, there are only two comparative studies.  In \cite{Ristea2022}, authors compare complex vs real spectrograms so as to discuss the advantages of additionally feeding the phase of the frequency-domain signal into the DL-model for epicenter localization. Similarly in \cite{Nakano2022}, a time vs frequency domain comparison using CNNs for long-duration waveforms ($>$150 sec.) is carried out. On the general discussion about whether a time or frequency-domain signal should be fed into a DL model, \cite{Nossier2020} studies various architectures that utilize audio signals. 

In order to answer these research questions, we design a series of comparative experiments. These experiments hinge on the selection of input representations, encompassing both the time and frequency domains, to unravel the influence of signal representation on our results. In addition, we introduce a large-scale strong motion record set, namely the AFAD-1218, a substantial collection comprising over 36,000 Turkish strong motion records. In order to investigate the influence of DL models, we employ two prevalent CNN architectures as encoders: the ResNet \cite{He2015} and the TCN \cite{TCN2016}. 

The remainder of the paper is as follows. Next, we introduce AFAD-1218 set, and how it is preprocessed for use in our DL models. Detailed descriptions of the DL models used in our experiments are presented in Section 3. Section 4 describes the experimental setup and discusses the experiment's findings. The final section concludes the paper and procures future directions.

\begin{figure*}[h]
    \centering
    \begin{subfigure}{0.22\columnwidth}	
        \includegraphics[trim=0 0 0 0,clip=true,width = 1\textwidth]{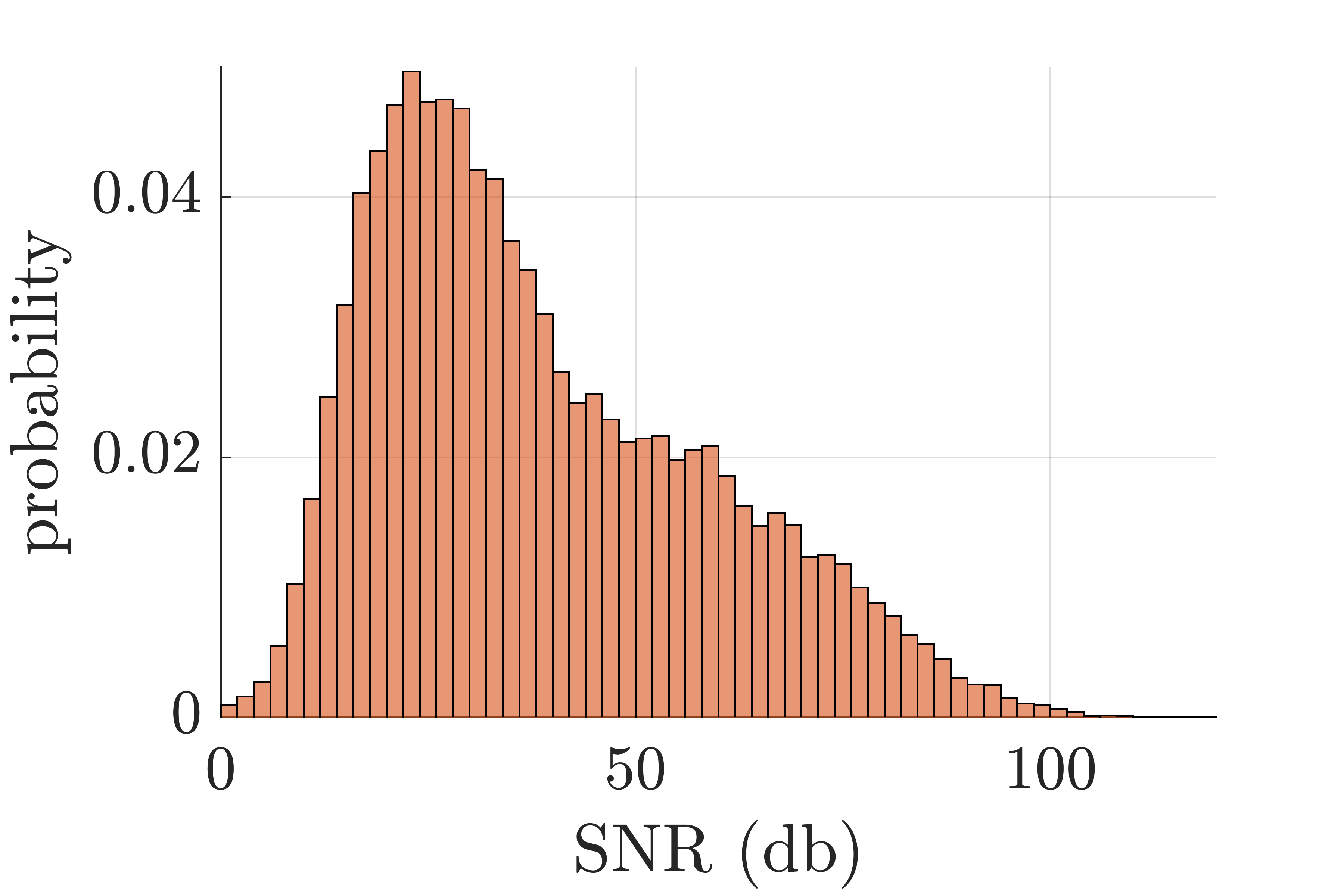}
        \caption{SNR (dB)}
    \end{subfigure}
    \begin{subfigure}{0.22\columnwidth}	
        \includegraphics[trim=0 0 0 0,clip=true,width = 1\textwidth]{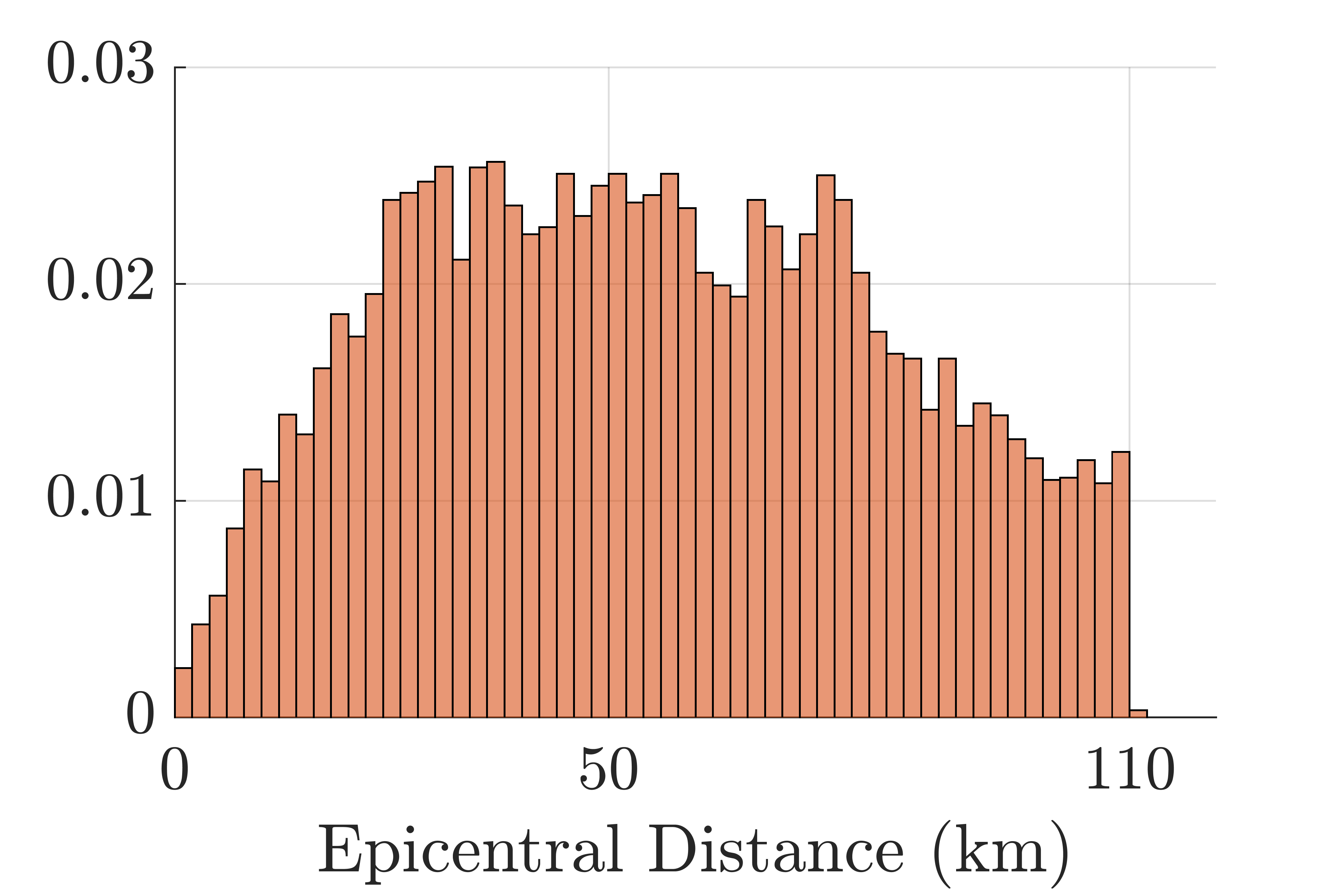}
        \caption{Epi. Dist. (km)}
    \end{subfigure}
    \begin{subfigure}{0.22\columnwidth}	
        \includegraphics[trim=0 0 0 0,clip=true,width = 1\textwidth]{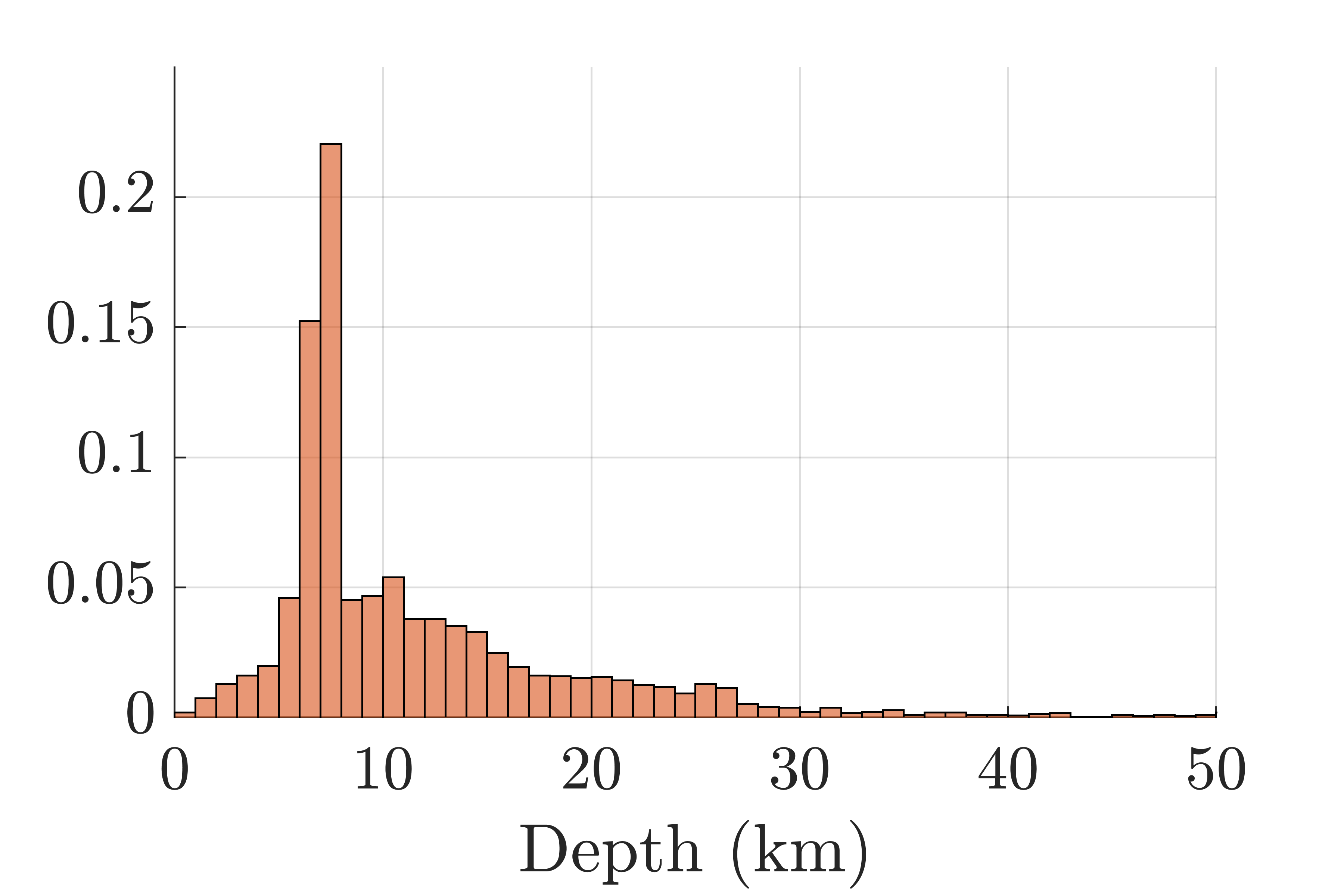}
        \caption{Depth (km)}
    \end{subfigure}
    \begin{subfigure}{0.22\columnwidth}	
        \includegraphics[trim=0 0 0 0,clip=true,width = 1\textwidth]{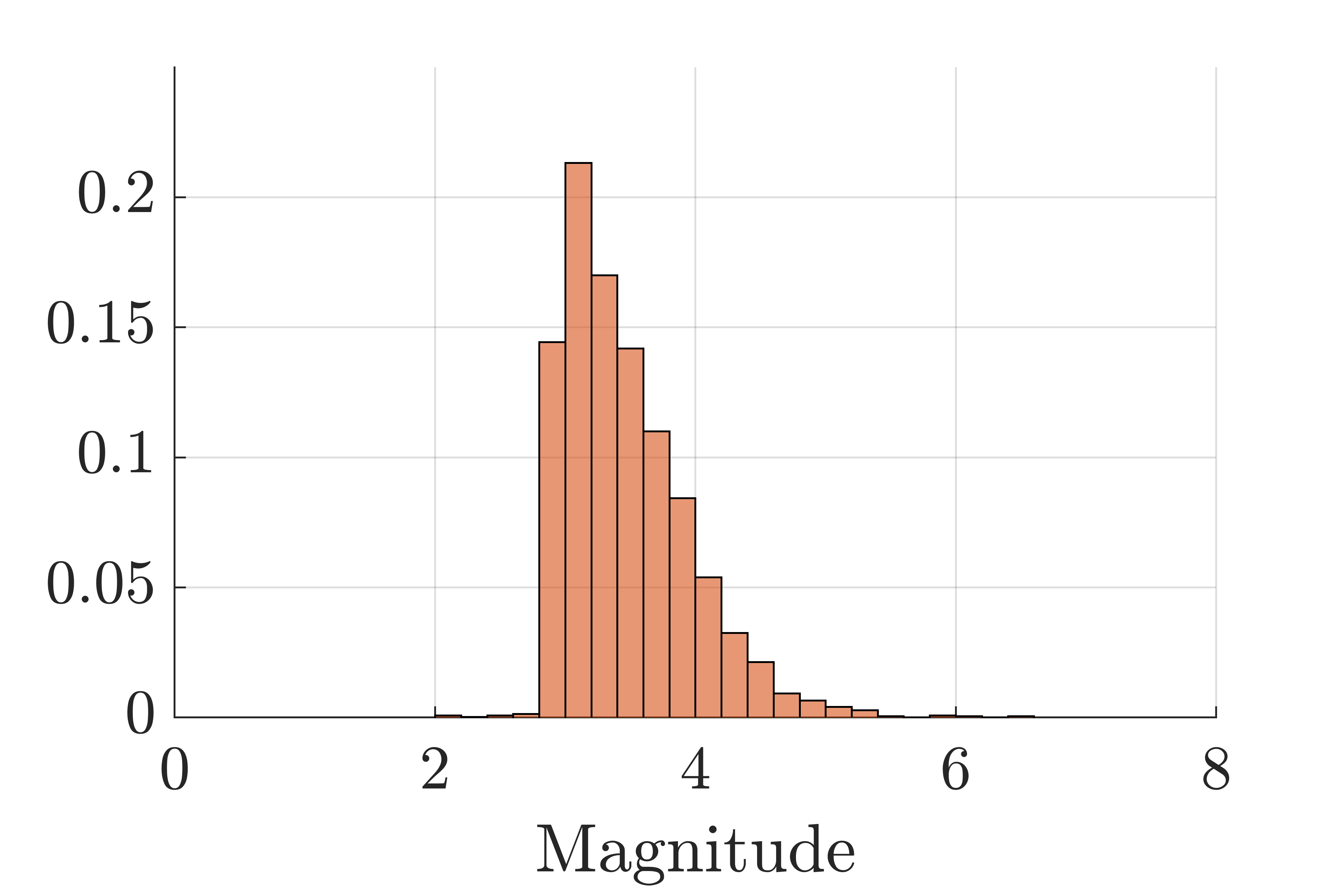}
        \caption{Magnitude}
    \end{subfigure}

    \centering
    \begin{subfigure}{0.220\columnwidth}	
        \includegraphics[trim=0 0 0 0,clip=true,width = 1\textwidth]{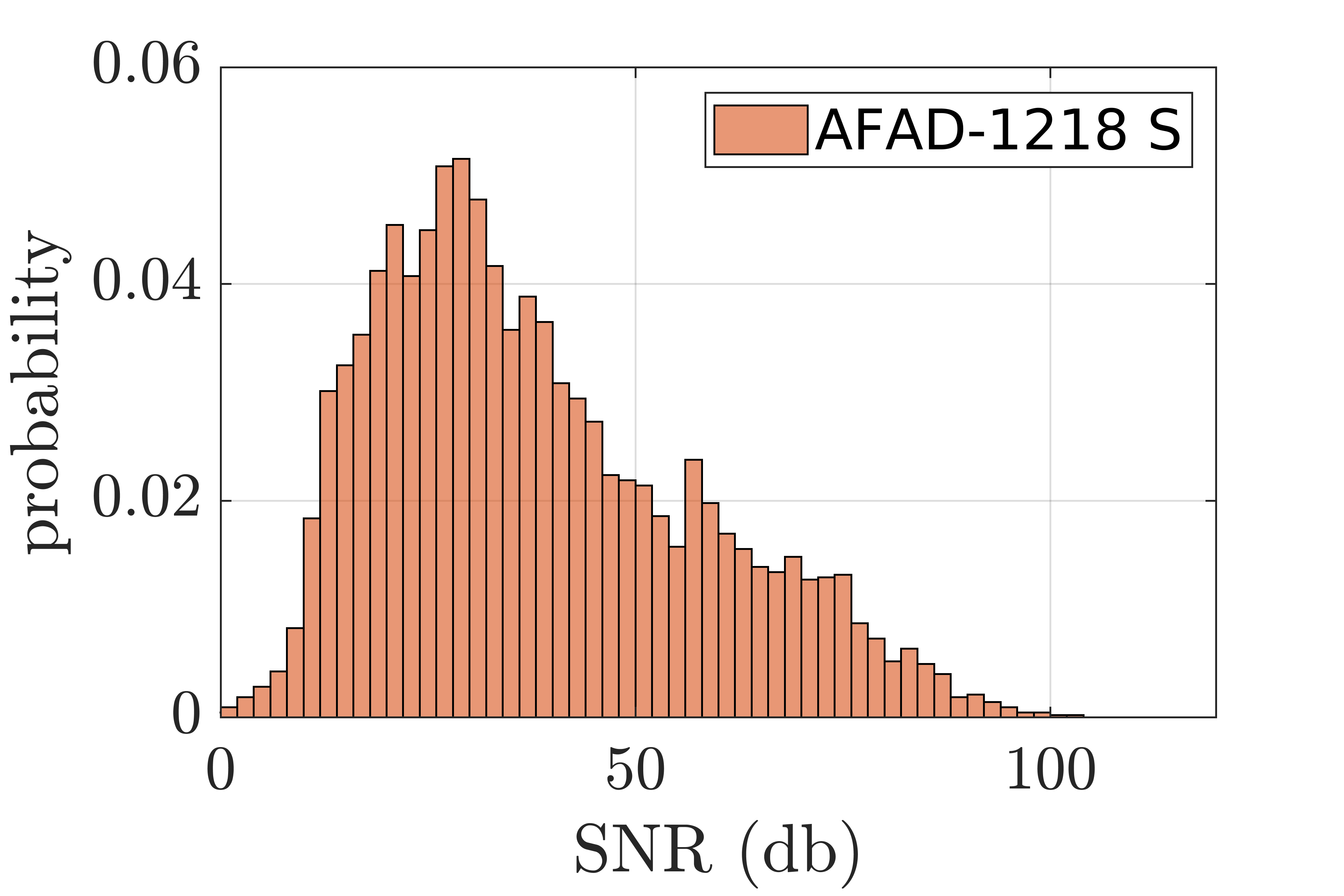}
        \caption{SNR (dB)}
    \end{subfigure}
    \begin{subfigure}{0.22\columnwidth}	
        \includegraphics[trim=0 0 0 0,clip=true,width = 1\textwidth]{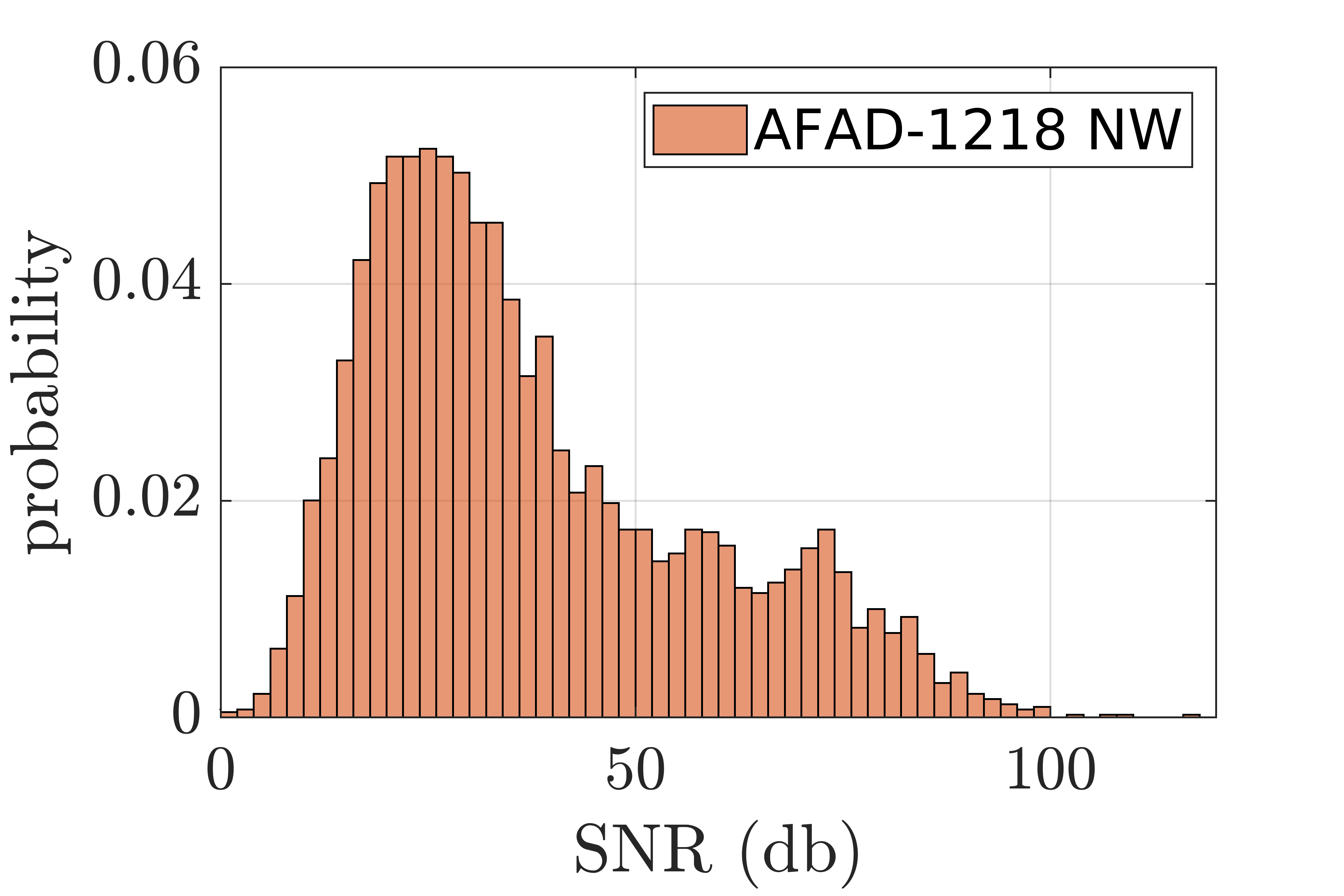}
        \caption{SNR (dB)}
    \end{subfigure}
    \begin{subfigure}{0.22\columnwidth}	
        \includegraphics[trim=0 0 0 0,clip=true,width = 1\textwidth]{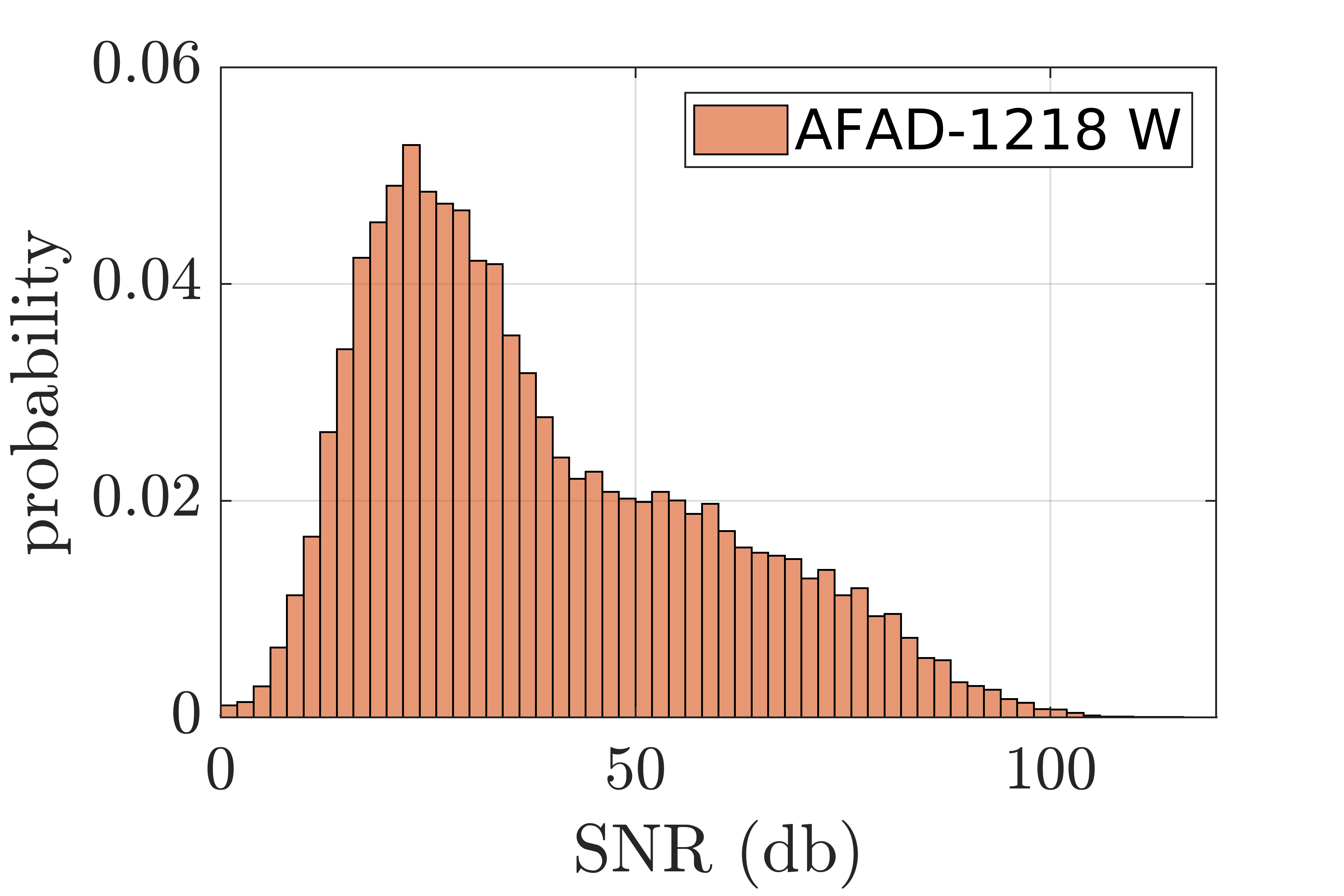}
        \caption{SNR (dB)}
    \end{subfigure}
    \begin{subfigure}{0.22\columnwidth}	
        \includegraphics[trim=0 0 0 0,clip=true,width = 1\textwidth]{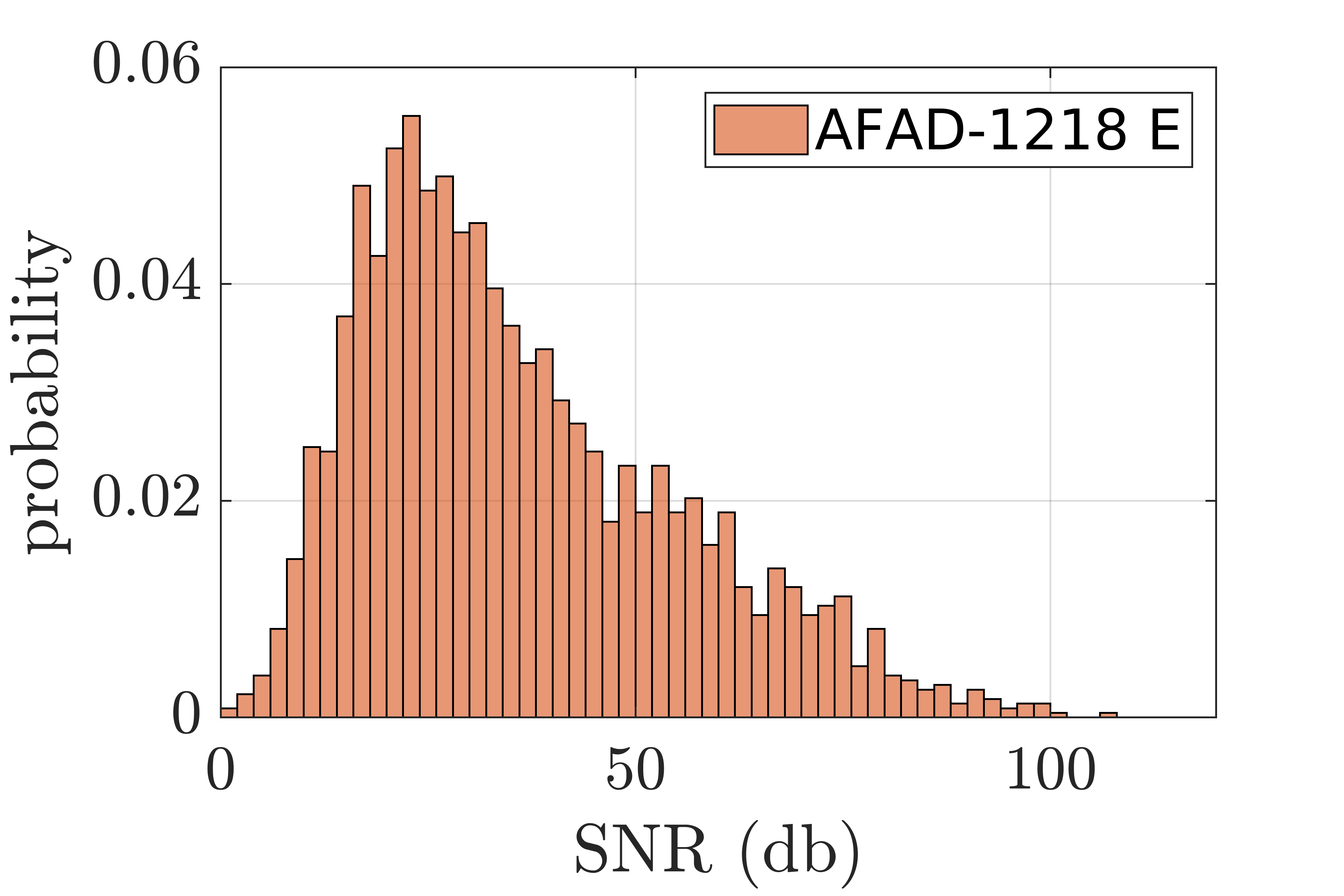}
        \caption{SNR (dB)}
    \end{subfigure}
    \caption{The distributions for the different properties of the AFAD-1218 set, along with the regional SNR distributions are depicted.}
    \label{fig:snr-epidist}
\end{figure*}

\section{AFAD-1218} % \& KRDAE-2022 Datasets}
The raw strong motion waveforms, i.e. waveforms for earthquake events having a magnitude larger than 3.5, collected for this study are publicly available and presented by AFAD \cite{AFAD}. At the time of the construction of the dataset, AFAD had 799 strong motion stations across Turkey that are located in different regions. Strong motion events collected at these stations since the 1990s are included in the public dataset. For every record, it is possible to access the time, epicenter coordinates, depth, magnitude, station number recording the earthquake, the coordinate of that station, and three-channel accelerometer waveforms recorded for the whole event in all directions. Since 2012, AFAD's recording sensors have been converted to digital, giving them a much wider dynamic range and higher data quality than analog stations \cite{Boore2005}.

In this study, we utilize {36,418} three-channel waveforms that belong to {3,655} separate events collected by 718 of AFAD's stations for 8 years, between 02.01.2012 and 19.12.2018. We name this dataset \textbf{AFAD-1218} and utilize our experiments on this set.  In Figure \ref{fig:AFAD-EQs}, all {3,655} events (red dots) and the station coordinated (green triangles) included in AFAD-1218 dataset are shown on a map of Turkey. 

There is wide variation in the duration of events in the AFAD-1218 dataset, which ranges from 5 to 300 seconds. We use only a short duration ($d$ = 15, 30 or 60 seconds) interval of the entire waveform to locate the epicenter. All records are in 100Hz sampling rate, hence the input signals either have 1500, 3000 or 6000 sample points.

In Figure \ref{fig:snr-epidist}, the statistical properties of the utilized AFAD-1218 dataset is provided. As seen from the figure, the set contains signals of varying SNR values as low as a few dBs and as high as 100dB. In addition, the SNR distributions of the four regions containing spatially extensive fault planes in Turkey, represented in Figures from  \ref{fig:snr-epidist}e to \ref{fig:snr-epidist}h, show no distinct dissimilarities. In Figure \ref{fig:snr-epidist}b, the probability distribution of the epicentral distances for each set is depicted. Following the general approach in the literature (as in the study by \citeA{Mousavi2020a}), in our experiments we also exclude waveforms that were received at epicentral distances greater than 110km, leaving us with the 27,185 waveforms of the entire AFAD-1218 set. The event-wise properties of the set are depicted in Figures \ref{fig:snr-epidist}c and \ref{fig:snr-epidist}d. AFAD-1218 contains events of diverse depths and magnitudes larger than 3.

\subsection{Peak Ground Acceleration Feature}
An accelerogram's Peak Ground Acceleration (PGA) represents the largest absolute acceleration measured at a specific site during an earthquake, calculated from the raw waveform for each direction separately. For this purpose, we define \(t_{PGA}\) as the median value of the PGA time instants computed for each channel separately. %(see Eq. \ref{Eq:tPGA}).
Finally by segmenting the $d$-seconds waveform centered around the calculated \(t_{PGA}\), the input to our DL model is obtained. {We chose PGA as the reference point because it can be easily calculated from a signal without the need for developing a model, and it is a fundamental parameter that is related to the intensity of the earthquake and represents the region where the signal is strongest, thus maximizing the likelihood of finding the most informative part of the data for strong motion. While there are studies that attack the same problem (i.e. epicenter localization) and select the P-wave arrival as the anchor point \cite{Mousavi2020a}, for the the aforementioned reasons, we focused on PGA here. It is worth mentioning that although PGA might not be the primary reference point chosen for earthquake localization studies, the higher earthquake intensity and improved signal-to-noise ratio around this parameter allow for potentially useful information around its vicinity. This setup also provides an opportunity to test our hypothesis regarding extracting valuable information from regions of the signal with increased signal clarity and significance in motion analysis.}
%Although using PGA as a reference may not always be the most accurate choice, we are also assessing the impact of whether the deep learning model is roughly accurate in the relevant area, thus measuring its robustness in this regard.

%\begin{equation}
%t_{PGA} =
%median(t^{N-S}_{PGA}, t^{E-W}_{PGA}, t^{U-D}_{PGA})
%\label{Eq:tPGA}
%\end{equation}

In our experiments, the short-duration signals are collected around the \(t_{PGA}\) of a given waveform. For example, for a $d$ seconds signal ($d$ having values of [15, 30, 60] seconds in this paper), the region between \(t_{PGA}\)-$\frac{d}{2}$ to \(t_{PGA}\)+$\frac{d} {2}$ seconds is selected as input. For cases, the \(t_{PGA}\) is too early or too late in the signal (i.e. \(t_{PGA}\)-$\frac{d}{2}<0$ or \(t_{PGA}\)+$\frac{d}{2}>t_{total}$), the first or the last $d$ seconds of the waveform is selected as input. %Which part of the signal carries the required information is an important open problem in the field. {The rationale behind conducting experiments with varying length regions around the PGA stems from the notion that the PGA indicates the segment of a ground motion with highest shaking intensity. Therefore, it is plausible that the essential information required to identify the underlying source structure and geology may reside around this segment of the motion.}

\begin{figure}[t]
    \centering
    \includegraphics[trim=18 0 25 15,clip=true,width = 0.8\textwidth]{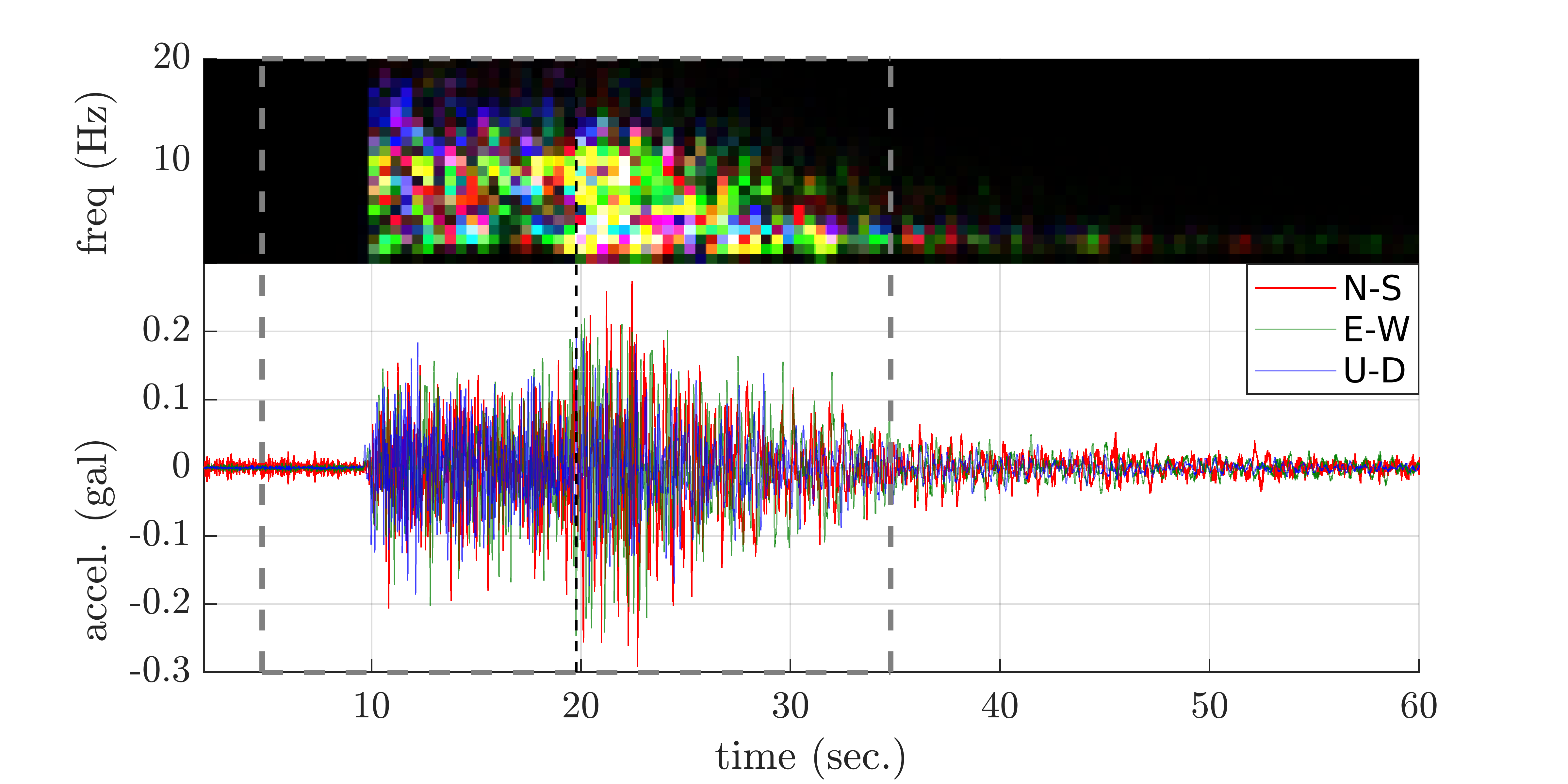}
    \caption{Three-channel waveform and the corresponding three-channel energy spectrogram are depicted. The short duration signal of $d$=30 seconds around \(t_{PGA}\)=19.78 sec. is shown with dashed lines. The waveform is recorded by the AFAD sensor at Van Gevaş Meteorology Directorate. This is a 4.2 M$_w$ earthquake that took place on 23 May 2018, at 22:10:43 local time, near İçlikaval region, Bitlis, Turkey. }
    \label{fig:spectro}
\end{figure}

\subsection{Spectrogram Calculation}

As previously mentioned in the Introduction, we conduct
comparative analyses by using time and frequency domains as inputs in our experiments. For this purpose, the magnitude squared of the Short-Time Fourier Transform (STFT) $s[n]$, namely the magnitude spectrogram of the $d$ seconds signal is created.

The spectrogram $s_i[n]$ is a $d$ seconds long sequence with three channels, where $i$ corresponds to each one of the orthogonal directions of the recorded event. For each direction $i$, the spectrogram is calculated separately.  The size of the window function $L$ used in STFT calculation is selected as 1 second (i.e. 100 sample points for 100Hz). Since the 1-second window function designates the size of the calculated Fourier transforms, for 100 samples in 100Hz, we obtained a frequency range of 0 to 50Hz (i.e. providing a $51$-long frequency resolution). The window function is propagated with a traditional $L/2$ = 0.5 s. hop, hence giving a time resolution of 2$d$-1 (i.e. $29$ for $d$=15 s. and $59$ for $d$= 30 s.). As a result, the input spectrograms are of size $51$ $\times$ (2$d$-1) $\times$ $3$. In Figure \ref{fig:spectro}, the entire spectrogram of a sample event is provided. In Figure \ref{fig:spectro}, the three directional channels of $s_i[n]$ are mapped to RGB false color (E-W as red, N-S as green, U-D as blue). The \(t_{PGA}\) instant and the $d$ = {30} seconds region around \(t_{PGA}\) are also depicted.

\section{The Methodology}
This section presents brief information on the utilized models and elucidates the training and parameter optimization phases.

\subsection{The models} 
\begin{figure*}[t]
\centering
    \includegraphics[trim=0 0 0 0,clip=true,width=0.9\textwidth]{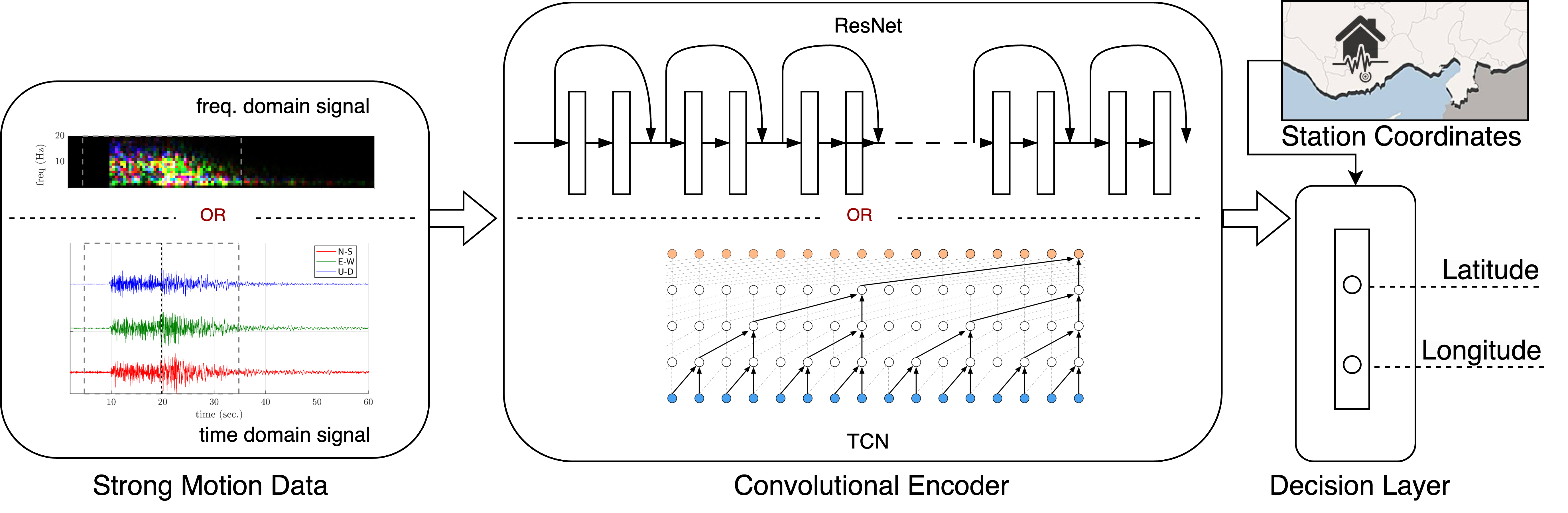}
\caption{The architectures of the implemented deep residual and TCN-based networks }
\label{fig:model-arch}
\end{figure*}

In order to utilize the strong motion records represented in either the time or the frequency domain, we propose two neural network architectures for operation.
Both architectures operate in the form outlined in Figure \ref{fig:model-arch}. The models are comprised of an encoder, followed by a decision layer that outputs the latitude and longitude (i.e. epicenter) of the event. As the encoder, we employ either deep residual network or the temporal convolutional networks (TCN) to find the best architecture of choice for capturing seismic features out of strong motion records.

 In addition to the seismic features extracted from the strong motion records via the convolutional encoders, the station latitude and longitude are concatenated to the first decision layer (Figure \ref{fig:model-arch}). By doing so, we expect the model to associate its input signals to source regions and deduce the arrival direction, allowing the model to better elucidate the local generation mechanisms rather than just identifying the presence of an event. 
 
Fully-connected decision layers reshape this input into two regression nodes, which predict the relative latitude and the longitude of the epicenter. The decision layer is chosen as a simple fully-connected layer that consists of two output neurons and Rectified Linear Unit (ReLU) activations for comparability against the existing literature such as \cite{Perol2018, Lomax2019, Kriegerowski2019}. By ``relative'', we mean the relative location of the epicenter with respect to the recording station location. For this purpose, in both architectures, the ground truth $\textbf{g}$ is not the actual epicenter coordinates, but the difference between the epicenter location and the station coordinates (Eq. \ref{gtruth}).

\begin{equation}
\textbf{g} = \left[ Lat_{epi}\quad Lon_{epi}\right] - \left[ Lat_{station}\quad Lon_{station}\right]
\label{gtruth}
\end{equation}

The following subsections provide a detailed description of the utilized convolutional architectures.

\subsubsection{Deep Residual Networks}

The first encoder we utilize is based on the ResNet architecture \cite{He2015}, which has been applied to seismic waveforms in \cite{Ristea2022}. We extend the architecture and propose a new layer model designed to extract features from recordings of a single station.  %Workflow of the proposed model is outlined in FigXXX.

The encoder consists of convolutional and residual blocks. The convolutional block processes a dx51x3 dimensional volume of input when operating in the frequency domain, or dxsx1 size of input when operating in the time domain by 32 convolving kernels of height to width ratio of 7/3 with stride 1, and zero padding with [3,1] for the same convolution, followed by a max pooling layer with spatial extent of 3 and stride [1,2]. 

The two convolutional layers are followed by three residual blocks, each of which consists of three convolutional layers stacked in tandem and a max pooling layer to achieve richer hierarchical representations. Each residual block starts with 1x1 convolution operation to adapt depths without losing spatial information, uses batch normalization before pooling operation, and contains a skip connection to the input of that layer and passes it to two layers ahead, allowing the network to reuse information throughout the architecture. The kernel sizes range from 3x3 to 7x3, and convolutional layers are stacked to grow the receptive field while keeping the model small. 

\subsubsection{Temporal Convolutional Networks}

Strong motion records are by definition time series data and hence have a sequential nature. In the literature, processing sequential data, instead of a spatial signal such as an image, is carried out with a special family of convolutional networks, namely the Temporal Convolutional Networks (TCN) \cite{TCN2016}. TCNs are a part of the autoregressive feedforward model family that utilizes causal dilated convolutions and residual connections, and are highly effective for sequential modeling. %They have a wide range of applications, including speech recognition, time series analysis, natural language processing, and signal processing. 
Their distinct features allow for the efficient management of time-related relationships.

In their recent work, \cite{Mousavi2019} utilize TCNs to process broadband seismic waveforms for epicentral distance prediction.  
The original model, which we also utilize for our time domain experiments, receives a three-channel time signal with a dimension of 6000x3, where each minute encompasses 100 samples per second. We modify this architecture and apply it also to frequency domain signals. For this purpose, we append max-pooling layers to the end of the first three skip connections in the TCN architecture.

\subsection{Two-phase learning}

The geological diversity across Turkey results in varying seismicity around the county. To address this, we suggest a hybrid, two-stage training approach that employs ResNet and TCN-based encoders as feature-sharing layers, each in its own stage. In the first phase, the entire data set is fed to the base architecture to capture generalized representations across Turkey. In the second phase, we leverage the pre-trained model to transfer the features acquired in the first phase to four high seismic hazard regions of Turkey: Eastern, Western, Southern, and North-Western, each of which exhibits distinct event characteristics. {The four regions were selected based on SNR of the recordings, faults and distribution of the earthquakes in the dataset}. By freezing the weights of the decision layer and allowing only the FC layers to be optimized, we anticipate the model which initially trained on data from the entire country to fine-tune itself to capture the seismic pattern and regimes of these high-activity regions.
 
We split the data into training and test sets, with the training set comprising historical data from January 2012 to August 2017, and the test set assessing model performance using subsequent data. By doing this, we aim to enable the system to use historical data for predicting future event locations and to make it applicable in real-world scenarios, we employ a cross-validation approach that maintains the chronological order of events. This ensures that the model reflects real-world performance and reveals any time-dependent features.

\subsection{Parameter optimization}

For parameter optimization, each model trained on datasets of 15, 30, or 60-second signals, using a convolution operator applied to batches of 64 inputs. Dropout was applied to the final decision layers at a rate and 50\% to improve generalization capability. We utilized the Adam optimizer, starting from 1e-5 and reducing the learning rate by 0.9 every [5, 10, 20] epochs for [100, 200, 300] epochs. Learning was based on either Mean Squared Error (MSE) or Mean Absolute Error (MAE) to assess the impact of the chosen loss metric.

\section{Experiments and Results}
In our experiments, we compare and present the relative functionalities of the proposed models in a number of facets: such as the input representation domain as time or frequency, input signal duration anchored around the PGA value, network architecture and SNR value of a given input signal. 
For this purpose, we train 120 different models on input data of 15, 30, and 60 seconds in length, separately in frequency and time domains, utilizing both the ResNet and TCN architectures. This training is conducted for the entire AFAD-1218 dataset and four subsets corresponding to four seismic regions of Turkey. Each of the 120 experiments includes its own hyperparameter search, totaling approximately 500 different experiments and 5000 GPU hours.%Table \ref{tab:transfer_results} tabulates and ranks (from left to right) the best-performed models under the same hyperparametric alignment. 

\begin{table*}[h]
\centering
\caption{Model performances of experiments using the entire AFAD-1218 set, with and without the exclusion of SNR values
below 25 dB.(results are in km)}
\label{tab:turkey_results}
%\resizebox{\textwidth}{!}{%
\begin{tabular}{|cc|cc|cc|}
\hline
\multicolumn{2}{|c|}{\textbf{}}             & \multicolumn{2}{c|}{\textbf{SNR \textgreater 0}} & \multicolumn{2}{c|}{\textbf{SNR \textgreater 25}} \\ \hline
\multicolumn{1}{|c|}{Domain} & Duration (s) & \multicolumn{1}{c|}{ResNet}    & TCN   & \multicolumn{1}{c|}{ResNet}    & TCN    \\ \hline
\multicolumn{1}{|c|}{Time}        & 15   & \multicolumn{1}{c|}{65.26} & 65.77 & \multicolumn{1}{c|}{54.40} & 53.56 \\ \hline
\multicolumn{1}{|c|}{Frequency}   & 15   & \multicolumn{1}{c|}{71.73} & 65.50 & \multicolumn{1}{c|}{60.74} & 53.03 \\ \hline
\multicolumn{1}{|c|}{Time}        & 30   & \multicolumn{1}{c|}{66.45} & 66.11 & \multicolumn{1}{c|}{53.31} & 53.77 \\ \hline
\multicolumn{1}{|c|}{Frequency}   & 30   & \multicolumn{1}{c|}{75.44} & 65.43 & \multicolumn{1}{c|}{59.53} & 53.68 \\ \hline
\multicolumn{1}{|c|}{Time}        & 60   & \multicolumn{1}{c|}{71.57} & 68.53 & \multicolumn{1}{c|}{58.33} & 58.00 \\ \hline
\multicolumn{1}{|c|}{Frequency}   & 60   & \multicolumn{1}{c|}{74.45} & 68.20 & \multicolumn{1}{c|}{62.16} & 57.94 \\ \hline
\multicolumn{2}{|c|}{Mean}               & \multicolumn{1}{c|}{70.8}  & 66.6  & \multicolumn{1}{c|}{58.1}  & 55.0  \\ \hline
\multicolumn{2}{|c|}{Standard Deviation} & \multicolumn{1}{c|}{3.8}   & 1.3   & \multicolumn{1}{c|}{3.2}   & 2.1   \\ \hline
\end{tabular}%
%}
\end{table*}

The results of the experiments trained using the entire set are shown in Table \ref{tab:turkey_results}. In this table, we see that the model performances tend to decline as the input duration increases (e.g., 60), possibly suggesting that models struggle to extract information beyond the most informative parts of the signals. Another significant finding is the clear impact of signal-to-noise ratio (SNR), with notably improved results when SNR exceeds 25 dB. 
In this initial phase of our experiments, we observe no significant difference in performance between the ResNet and TCN architectures.

\begin{table*}[h]
\centering
\caption{Mean test errors of transfer experiments without the exclusion of accelerograms with SNR values below 25 dB (results are in km).}
\label{tab:low_snr_transfer_results}
%\resizebox{\textwidth}{!}{%
\begin{tabular}{|c|cccccc|cccccc|}
\hline
Architecture &
  \multicolumn{6}{c|}{ResNet} &
  \multicolumn{6}{c|}{TCN} \\ \hline
Domain &
  \multicolumn{3}{c|}{Frequency} &
  \multicolumn{3}{c|}{Time} &
  \multicolumn{3}{c|}{Frequency} &
  \multicolumn{3}{c|}{Time} \\ \hline
Duration (s) &
  \multicolumn{1}{c|}{15} &
  \multicolumn{1}{c|}{30} &
  \multicolumn{1}{c|}{60} &
  \multicolumn{1}{c|}{15} &
  \multicolumn{1}{c|}{30} &
  60 &
  \multicolumn{1}{c|}{15} &
  \multicolumn{1}{c|}{30} &
  \multicolumn{1}{c|}{60} &
  \multicolumn{1}{c|}{15} &
  \multicolumn{1}{c|}{30} &
  60 \\ \hline
Eastern &
  \multicolumn{1}{c|}{52.82} &
  \multicolumn{1}{c|}{53.29} &
  \multicolumn{1}{c|}{56.96} &
  \multicolumn{1}{c|}{52.8} &
  \multicolumn{1}{c|}{55.25} &
  57.04 &
  \multicolumn{1}{c|}{53.15} &
  \multicolumn{1}{c|}{53.39} &
  \multicolumn{1}{c|}{55.93} &
  \multicolumn{1}{c|}{53.13} &
  \multicolumn{1}{c|}{53.47} &
  56.05 \\ \hline
Southern &
  \multicolumn{1}{c|}{75.82} &
  \multicolumn{1}{c|}{73.92} &
  \multicolumn{1}{c|}{75.22} &
  \multicolumn{1}{c|}{69.35} &
  \multicolumn{1}{c|}{66.25} &
  71.1 &
  \multicolumn{1}{c|}{66.75} &
  \multicolumn{1}{c|}{67.52} &
  \multicolumn{1}{c|}{72.17} &
  \multicolumn{1}{c|}{67.23} &
  \multicolumn{1}{c|}{67.79} &
  72.34 \\ \hline
North-Western &
  \multicolumn{1}{c|}{69.69} &
  \multicolumn{1}{c|}{69.98} &
  \multicolumn{1}{c|}{74.13} &
  \multicolumn{1}{c|}{64.33} &
  \multicolumn{1}{c|}{62.99} &
  63.3 &
  \multicolumn{1}{c|}{65.95} &
  \multicolumn{1}{c|}{65.71} &
  \multicolumn{1}{c|}{64.85} &
  \multicolumn{1}{c|}{66.02} &
  \multicolumn{1}{c|}{65.99} &
  65.32 \\ \hline
Western &
  \multicolumn{1}{c|}{75.57} &
  \multicolumn{1}{c|}{73.93} &
  \multicolumn{1}{c|}{73.22} &
  \multicolumn{1}{c|}{66.03} &
  \multicolumn{1}{c|}{67.84} &
  70.25 &
  \multicolumn{1}{c|}{65.13} &
  \multicolumn{1}{c|}{65.55} &
  \multicolumn{1}{c|}{67.41} &
  \multicolumn{1}{c|}{65.33} &
  \multicolumn{1}{c|}{65.97} &
  67.54 \\ \hline
\end{tabular}%
\label{tab:reg-SNR}
%}
\end{table*}

In the following phase, we proceed with experiments using regional samples by transferring the weights from the models trained using the entire set. We designate four regions in Turkey, roughly spanning Western, Eastern, Southern and North-Western parts of the country. The results for this phase of the experiments are displayed in Table \ref{tab:reg-SNR}. Similarly to Table \ref{tab:turkey_results}, shorter input durations tend to perform better. We note performance discrepancies, potentially arising from the intricate geological characteristics of each seismic region. For these experiments, we continue to include lower than 25 dB SNR samples while transferring model parameters to regional models.

\begin{table*}[h]
\centering
\caption{Mean test errors of transfer experiments exluding the data with the SNR values below 25 dB (results are in km).}
\label{tab:high_snr_transfer_results}
%\resizebox{\textwidth}{!}{%
\begin{tabular}{|c|cccccc|cccccc|}
\hline
Architecture &
  \multicolumn{6}{c|}{ResNet} &
  \multicolumn{6}{c|}{TCN} \\ \hline
Domain &
  \multicolumn{3}{c|}{Frequency} &
  \multicolumn{3}{c|}{Time} &
  \multicolumn{3}{c|}{Frequency} &
  \multicolumn{3}{c|}{Time} \\ \hline
Duration (s) &
  \multicolumn{1}{c|}{15} &
  \multicolumn{1}{c|}{30} &
  \multicolumn{1}{c|}{60} &
  \multicolumn{1}{c|}{15} &
  \multicolumn{1}{c|}{30} &
  60 &
  \multicolumn{1}{c|}{15} &
  \multicolumn{1}{c|}{30} &
  \multicolumn{1}{c|}{60} &
  \multicolumn{1}{c|}{15} &
  \multicolumn{1}{c|}{30} &
  60 \\ \hline
Eastern &
  \multicolumn{1}{c|}{53.39} &
  \multicolumn{1}{c|}{52.9} &
  \multicolumn{1}{c|}{54.65} &
  \multicolumn{1}{c|}{49.07} &
  \multicolumn{1}{c|}{50.71} &
  52.57 &
  \multicolumn{1}{c|}{49.55} &
  \multicolumn{1}{c|}{49.91} &
  \multicolumn{1}{c|}{52.98} &
  \multicolumn{1}{c|}{49.39} &
  \multicolumn{1}{c|}{50.19} &
  52.68 \\ \hline
Southern &
  \multicolumn{1}{c|}{67.36} &
  \multicolumn{1}{c|}{68.32} &
  \multicolumn{1}{c|}{72.53} &
  \multicolumn{1}{c|}{58.23} &
  \multicolumn{1}{c|}{56.39} &
  61.51 &
  \multicolumn{1}{c|}{60.12} &
  \multicolumn{1}{c|}{60.82} &
  \multicolumn{1}{c|}{66.18} &
  \multicolumn{1}{c|}{60.53} &
  \multicolumn{1}{c|}{61.12} &
  66.35 \\ \hline
North-Western &
  \multicolumn{1}{c|}{55.80} &
  \multicolumn{1}{c|}{52.19} &
  \multicolumn{1}{c|}{63.70} &
  \multicolumn{1}{c|}{47.73} &
  \multicolumn{1}{c|}{48.59} &
  50.76 &
  \multicolumn{1}{c|}{46.86} &
  \multicolumn{1}{c|}{47.46} &
  \multicolumn{1}{c|}{50.29} &
  \multicolumn{1}{c|}{47.11} &
  \multicolumn{1}{c|}{47.57} &
  50.38 \\ \hline
Western &
  \multicolumn{1}{c|}{59.97} &
  \multicolumn{1}{c|}{50.87} &
  \multicolumn{1}{c|}{61.98} &
  \multicolumn{1}{c|}{51.88} &
  \multicolumn{1}{c|}{53.86} &
  58.12 &
  \multicolumn{1}{c|}{51.76} &
  \multicolumn{1}{c|}{52.39} &
  \multicolumn{1}{c|}{56.14} &
  \multicolumn{1}{c|}{52.24} &
  \multicolumn{1}{c|}{52.39} &
  56.32 \\ \hline
\end{tabular}%
%}
\label{tab:reg-hiSNR}
\end{table*}

To assess the impact of SNR in the transfer-learning experiments, we replicate the previous experiment by excluding low SNR samples, specifically using samples with SNR values higher than 25. The results of the experiments are provided in Table \ref{tab:reg-hiSNR}. 
We observe that performance differences between regions are mitigated, and the impact of SNR persists in the transfer-learning experiments. 
Excluding low SNR records resulted in an 18\% reduction in mean prediction error in experiments utilizing the entire set and a 14.9\% reduction in the average of all transfer (i.e. regional) experiments.

\begin{figure*}[h]
    \centering
    \begin{subfigure}{0.22\columnwidth}	
        \includegraphics[trim=0 0 0 6,clip=true,width = 1\textwidth]{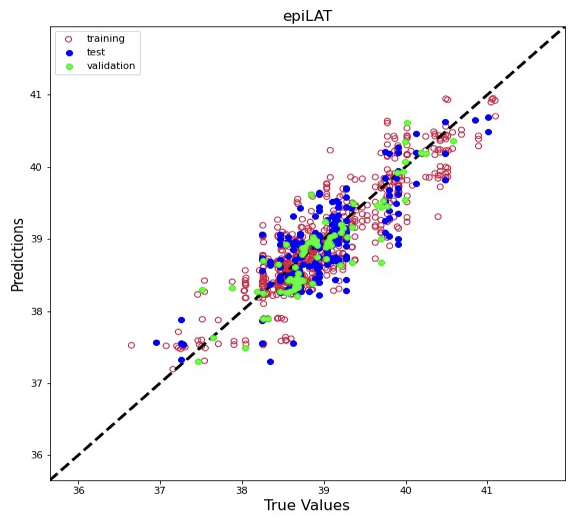}
        \caption{Latitude (E)}
    \end{subfigure}
    \begin{subfigure}{0.22\columnwidth}	
        \includegraphics[trim=0 0 0 7,clip=true,width = 1\textwidth]{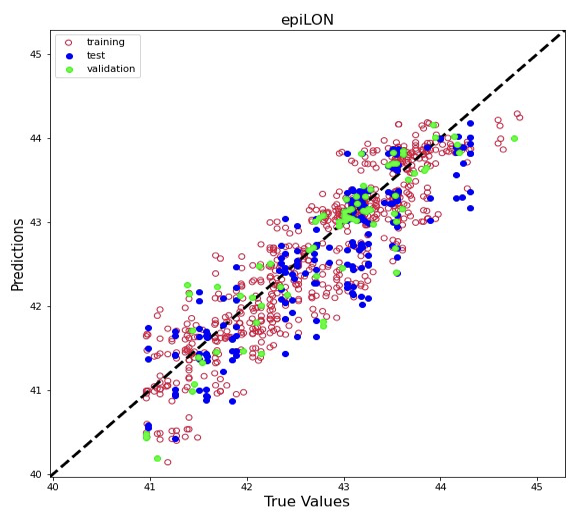}
        \caption{Longtitude (E)}
    \end{subfigure}
    \begin{subfigure}{0.22\columnwidth}	
        \includegraphics[trim=0 0 0 7,clip=true,width = 1\textwidth]{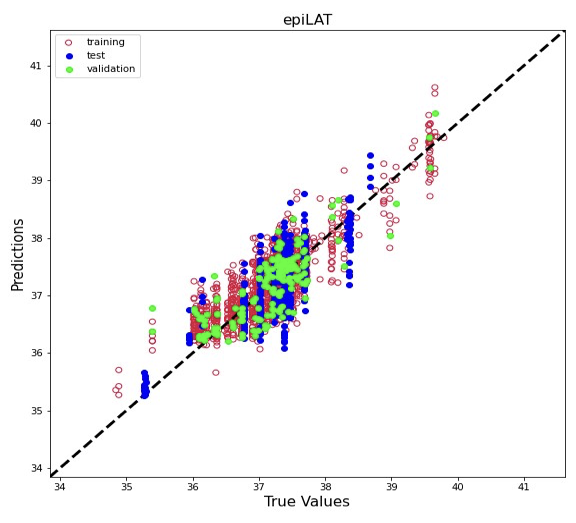}
        \caption{Latitude (S)}
    \end{subfigure}
    \begin{subfigure}{0.22\columnwidth}	
        \includegraphics[trim=0 0 0 6,clip=true,width = 1\textwidth]{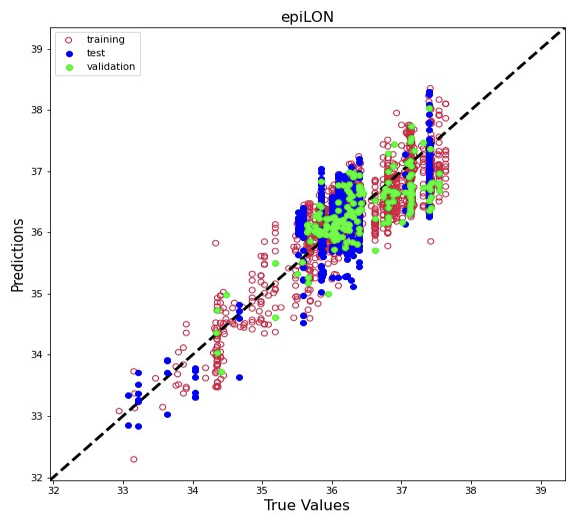}
        \caption{Longtitude (S)}
    \end{subfigure}

    \centering
    \begin{subfigure}{0.220\columnwidth}	
        \includegraphics[trim=0 0 0 7,clip=true,width = 1\textwidth]{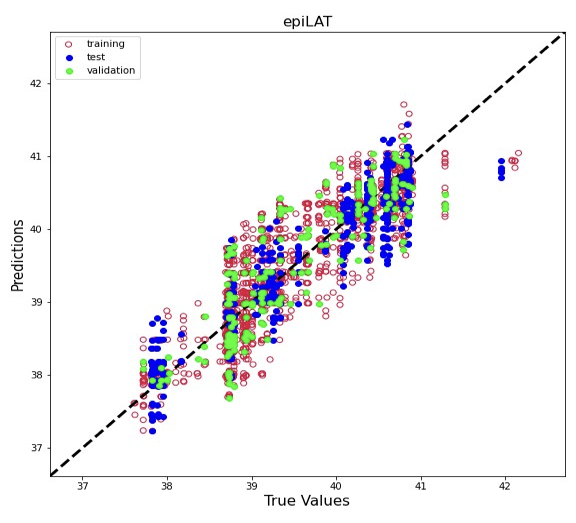}
        \caption{Latitude (NW)}
    \end{subfigure}
    \begin{subfigure}{0.22\columnwidth}	
        \includegraphics[trim=0 0 0 6,clip=true,width = 1\textwidth]{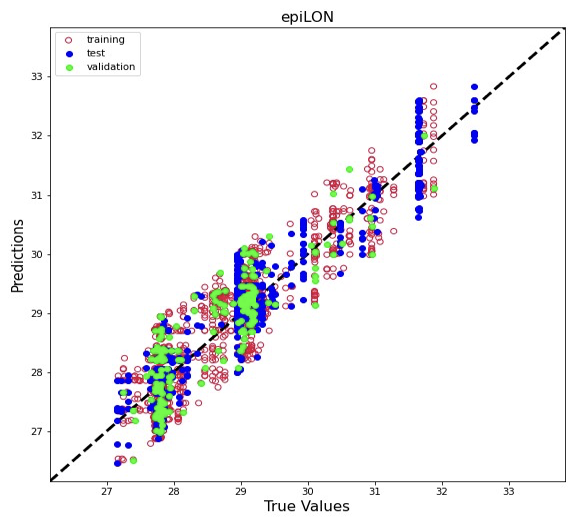}
        \caption{Longtitude (NW)}
    \end{subfigure}
    \begin{subfigure}{0.22\columnwidth}	
        \includegraphics[trim=0 0 0 7,clip=true,width = 1\textwidth]{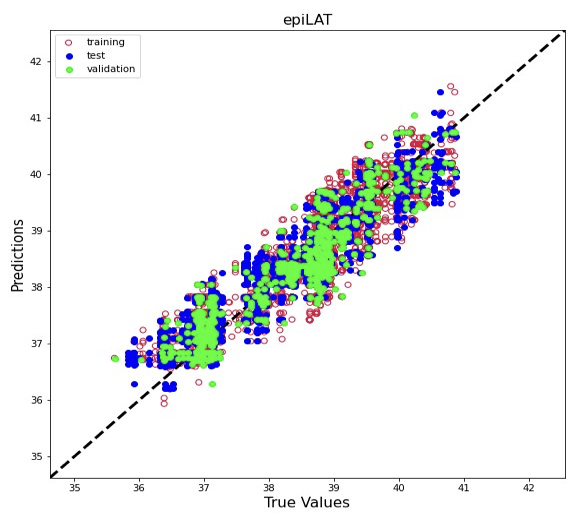}
        \caption{Latitude (W)}
    \end{subfigure}
    \begin{subfigure}{0.22\columnwidth}	
        \includegraphics[trim=0 0 0 6,clip=true,width = 1\textwidth]{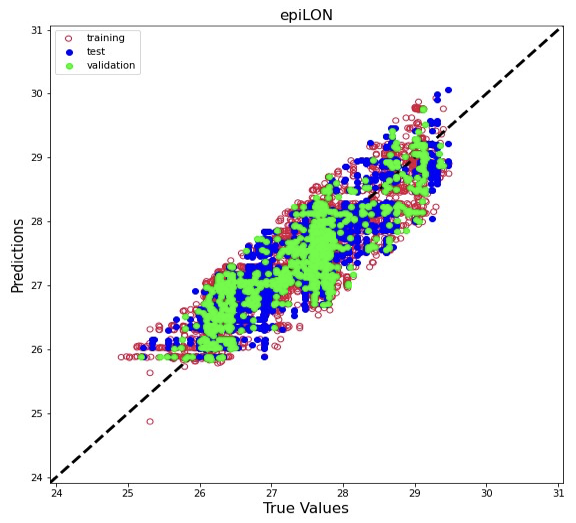}
        \caption{Longtitude (W)}
    \end{subfigure}
    \caption{Prediction results of the transfer learning experiments (E:Eastern, S:Southern, NW:NorthWestern and W:Western regions), including both low and high SNR samples.}
    \label{fig:prevstrue}
\end{figure*}

\begin{table*}[]
\small
\centering
\caption{Descriptive statistics for the mean test errors of the regional transfer learning experiments (results are in km).}
%\label{tab:transfer_results}
%\resizebox{\textwidth}{!}{%
\begin{tabular}{|c|c|cc|cc|cc|cc|}
\hline
{SNR \textgreater{}= 25} &
  Region &
  \multicolumn{2}{c|}{Eastern} &
  \multicolumn{2}{c|}{North-Western} &
  \multicolumn{2}{c|}{Western} &
  \multicolumn{2}{c|}{Southern} \\ \cline{2-10} 
 &
   &
  \multicolumn{1}{c|}{Mean} &
  Std &
  \multicolumn{1}{c|}{Mean} &
  Std &
  \multicolumn{1}{c|}{Mean} &
  Std &
  \multicolumn{1}{c|}{Mean} &
  Std \\ \cline{2-10} 
 &
  Overall &
  \multicolumn{1}{c|}{51.50} &
  1.81 &
  \multicolumn{1}{c|}{50.70} &
  4.64 &
  \multicolumn{1}{c|}{54.83} &
  3.48 &
  \multicolumn{1}{c|}{63.29} &
  4.56 \\ \hline
{Network} &
  ResNet &
  \multicolumn{1}{c|}{52.22} &
  1.83 &
  \multicolumn{1}{c|}{53.13} &
  5.40 &
  \multicolumn{1}{c|}{56.11} &
  4.16 &
  \multicolumn{1}{c|}{64.06} &
  5.77 \\ \cline{2-10} 
 &
  TCN &
  \multicolumn{1}{c|}{50.78} &
  1.47 &
  \multicolumn{1}{c|}{48.28} &
  1.47 &
  \multicolumn{1}{c|}{53.54} &
  1.91 &
  \multicolumn{1}{c|}{62.52} &
  2.67 \\ \hline
{Duration} &
  15 s &
  \multicolumn{1}{c|}{50.35} &
  1.76 &
  \multicolumn{1}{c|}{49.38} &
  3.72 &
  \multicolumn{1}{c|}{53.96} &
  3.47 &
  \multicolumn{1}{c|}{61.56} &
  3.46 \\ \cline{2-10} 
 &
  30 s &
  \multicolumn{1}{c|}{50.93} &
  1.17 &
  \multicolumn{1}{c|}{48.95} &
  1.92 &
  \multicolumn{1}{c|}{52.38} &
  1.06 &
  \multicolumn{1}{c|}{61.66} &
  4.28 \\ \cline{2-10} 
 &
  60 s &
  \multicolumn{1}{c|}{53.22} &
  0.84 &
  \multicolumn{1}{c|}{53.78} &
  5.73 &
  \multicolumn{1}{c|}{58.14} &
  2.35 &
  \multicolumn{1}{c|}{66.64} &
  3.91 \\ \hline
{Domain} &
  Frequency &
  \multicolumn{1}{c|}{52.23} &
  1.86 &
  \multicolumn{1}{c|}{52.72} &
  5.75 &
  \multicolumn{1}{c|}{55.52} &
  4.23 &
  \multicolumn{1}{c|}{65.89} &
  4.31 \\ \cline{2-10} 
 &
  Time &
  \multicolumn{1}{c|}{50.77} &
  1.42 &
  \multicolumn{1}{c|}{48.69} &
  1.40 &
  \multicolumn{1}{c|}{54.14} &
  2.33 &
  \multicolumn{1}{c|}{60.69} &
  3.09 \\ \hline
% \multirow{Data} &
%   Number of records &
%   \multicolumn{1}{c|}{} &
%    &
%   \multicolumn{1}{c|}{} &
%    &
%   \multicolumn{1}{c|}{} &
%    &
%   \multicolumn{1}{c|}{} &
%    \\ \cline{2-10} 
%  &
%   Epicentral Distance &
%   \multicolumn{1}{c|}{} &
%    &
%   \multicolumn{1}{c|}{} &
%    &
%   \multicolumn{1}{c|}{} &
%    &
%   \multicolumn{1}{c|}{} &
%    \\ \hline
\end{tabular}%
%}
\label{tab:regdetail}
\end{table*}

In Figure \ref{fig:prevstrue}, prediction results of the transfer learning experiments for each trianing (red), validation (green) and test (blue) sample is presented. In these figures all samples, with both low and high SNR are included. Paralel to Tables \ref{tab:reg-SNR} and \ref{tab:reg-hiSNR}, transfer learning experiments that belong to Eastern and Southern regions provide better prediction distributions. This further underscores the influence of regional geological and geophysical characteristics on our results.

\begin{figure*}[h]
    \centering
    \begin{subfigure}{0.48\textwidth}	
        \includegraphics[trim=0 0 0 0,clip=true,width=\linewidth]{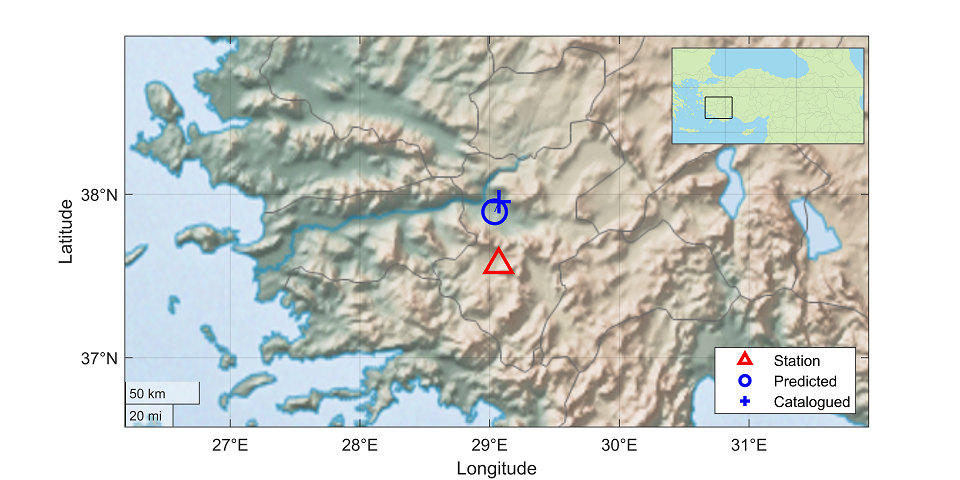}
        \caption{Station 2020 - Tavas, Denizli}
    \end{subfigure}
    \begin{subfigure}{0.48\textwidth}	
        \includegraphics[trim=0 0 0 0,clip=true,width=\linewidth]{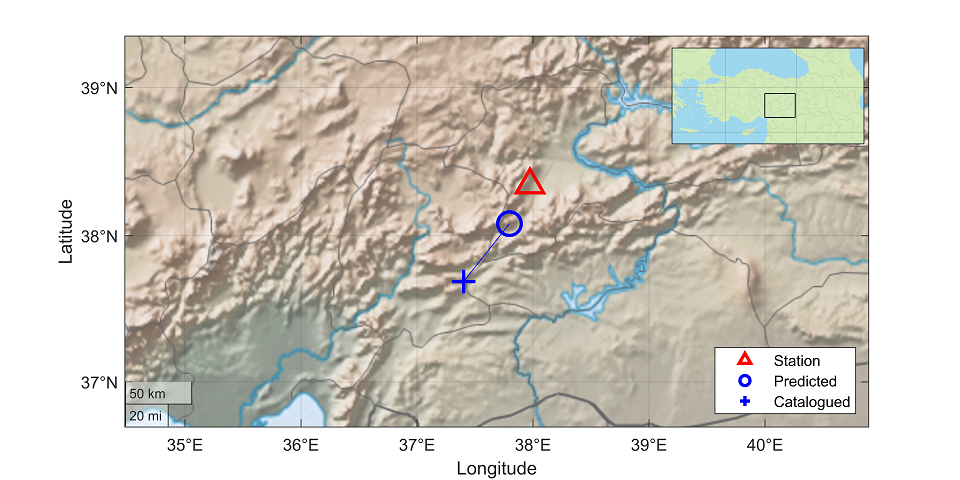}
        \caption{Station 4406 - Akçadağ, Malatya}
    \end{subfigure}
    \begin{subfigure}{0.48\textwidth}	
        \includegraphics[trim=0 0 0 0,clip=true,width=\linewidth]{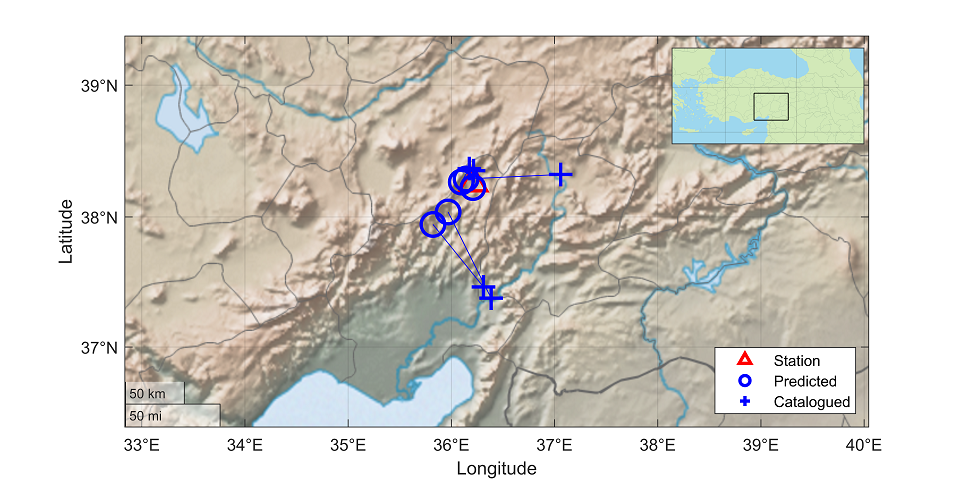}
        \caption{Station 129 - Tufanbeyli, Adana}
    \end{subfigure}
    \begin{subfigure}{0.48\textwidth}	
        \includegraphics[trim=0 0 0 0,clip=true,width=\linewidth]{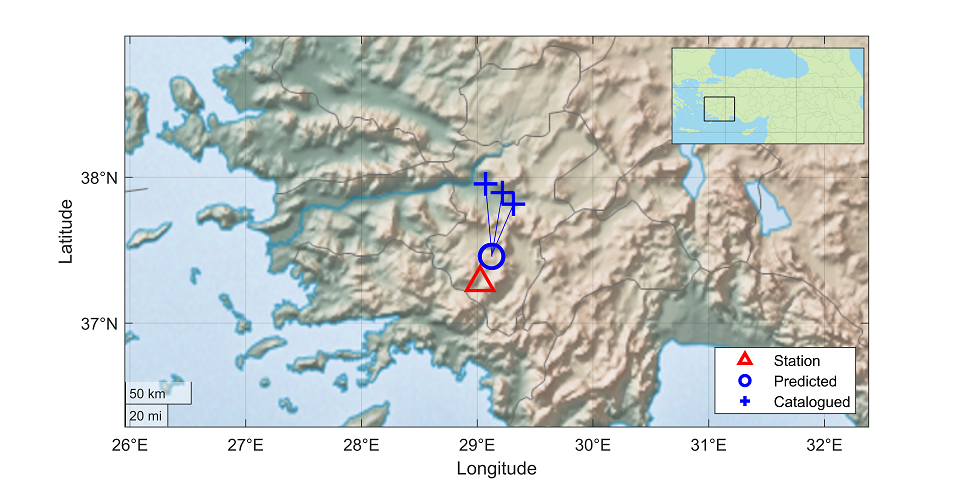}
        \caption{Station 2026 - Dereağzı, Denizli}
    \end{subfigure}
    \begin{subfigure}{0.48\textwidth}	
        \includegraphics[trim=0 0 0 0,clip=true,width=\linewidth]{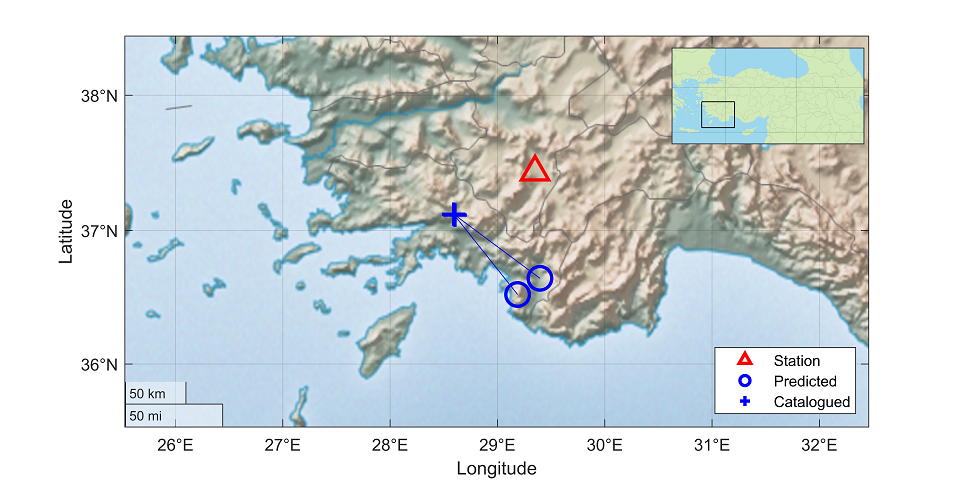}
        \caption{Station 2017 - Acıpayam, Denizli}
    \end{subfigure}
    \begin{subfigure}{0.48\textwidth}	
        \includegraphics[trim=0 0 0 0,clip=true,width=\linewidth]{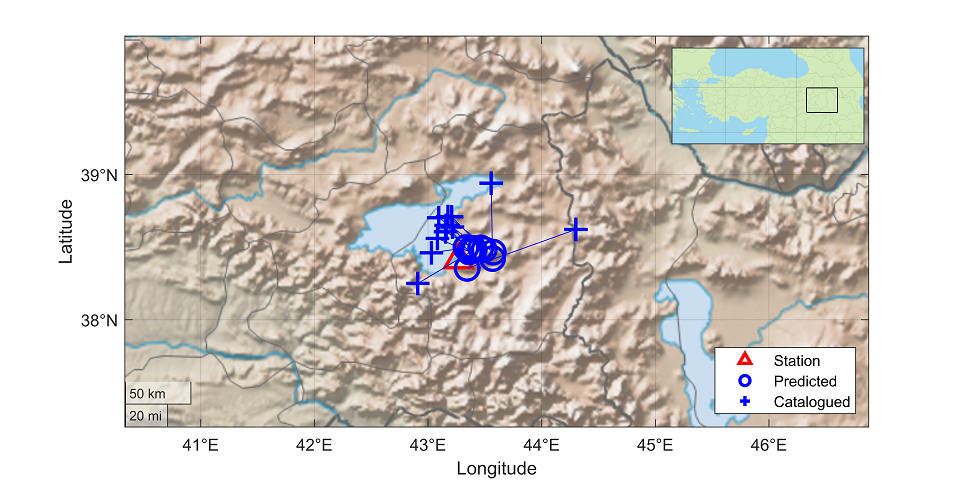}
        \caption{Station 6513 - Edremit, Van}
    \end{subfigure}
    \begin{subfigure}{0.48\textwidth}	
        \includegraphics[trim=0 0 0 0,clip=true,width=\linewidth]{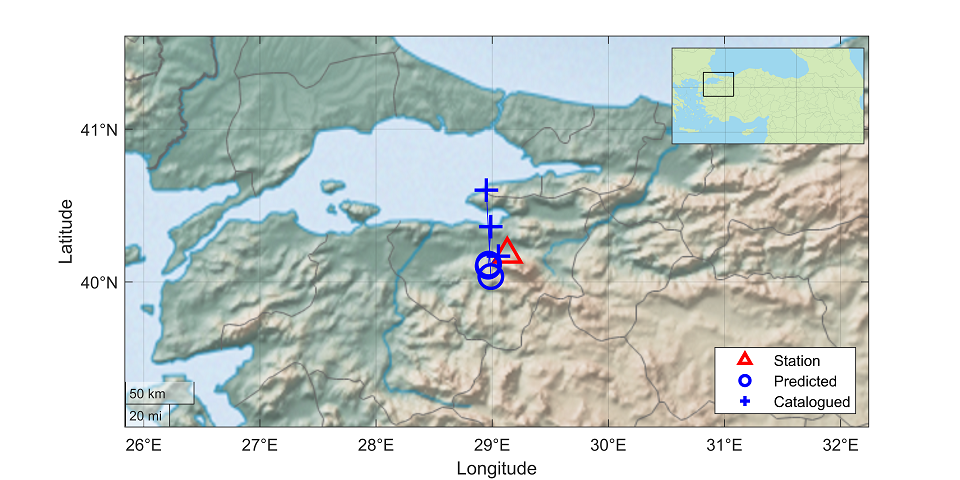}
        \label{fig:Station1620_lowRes}
        \caption{Station 1620 - Yıldırım, Bursa}
    \end{subfigure}
    \begin{subfigure}{0.48\textwidth}	
        \includegraphics[trim=0 0 0 0,clip=true,width=\linewidth]{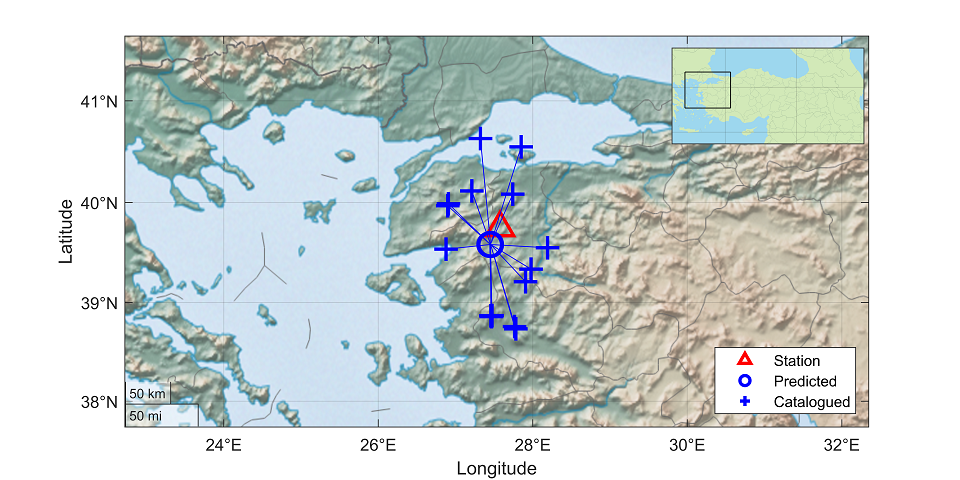}
        \caption{Station 1021 - Balya, Balıkesir}        
    \end{subfigure}
    \caption{Location estimates of transfer experiments. Each circle represents the estimated location. Triangles in red denote the station locations and the marked cross symbols (+) represent the actual epicentral locations.}
    \label{fig:comparison_figures}
\end{figure*}

For an overall assessment of all results, Table \ref{tab:regdetail} presents the averages of all transfer learning experiments for each region in the "Overall" row (third row). All experiments in this table are conducted using records with SNR values higher than 25 dB and with decision layers having a dropout rate of 0.5 (similarly to previous experiments). Subsequent rows of this table present average errors for specific parameters, again individually for each region. {For instance, the "ResNet" row (fourth row) represents the average results for all experiments that utilize this convolutional encoder. It is evident that the 15-seconds and 30-seconds regions centered around the PGA value yields superior results, consistent with our findings.} There is a slight performance preference for time-domain experiments. Regarding network architecture, TCN performs slightly better, in consistent with the results reported in \cite{Caglar2024}. Performance results indicate that predictions for Eastern and North-Western provinces have lower tendency to deviate from its observations then Western and Southern regions. The common ground between the two seismic zones is the station coverage. From the results, we observe that regions of spatially homogenous seismic activity requires either more records or additional directional information for the model to match the variability of waveforms and divulge any particular relevance to the earthquake sources. For example, the stations in Eastern and North-Western regions are aligned in an eastward and a northeastward direction, respectively. This implicitly gives a sense of arrival direction to the model as exemplified in Figure \ref{fig:comparison_figures}, in which all the predictions are directed towards the large fault at southwest of Station 1620. Results for different stations are presented in Figure \ref{fig:comparison_figures}, where predicted and catalouged epicenter locations with the station coordinates are depicted on map. We observe that for certain regions (Figure \ref{fig:comparison_figures}a), the prediction errors are minimal, typically within a few kilometers. However, for other regions, the results exhibit varying levels of accuracy.

\section{Conclusions and Future Directions}
In this paper, we aim to locate the epicenter of an earthquake in world coordinates using strong motion records collected at a single station. We address whether strong motion records from a single station contain earthquake location information and whether deep learning models can extract this information effectively. On different input signal durations, convolutional encoders, and signal-to-noise ratio (SNR) levels, we carry out several experiments. We introduce a large-scale strong motion dataset comprising over 36,000 three-channel records collected from Turkish earthquake stations and conduct both nationwide and regional experiments.

From our experiments, we draw significant conclusions regarding the necessary SNR level for epicenter localization and the impact of input signal duration. We find that focusing only on the strongest part of the signal yields crucial information for pinpointing the epicenter. Additionally, we note that regional seismic characteristics play a crucial role in model performance, indicating the feasibility of deploying such models in specific regions.

However, we also encounter challenges in training such models. While some individual results show precise localization within a radius of 5 km, the average success of the models is lower than anticipated. This is consistent with recent studies such as \cite{Caglar2024}, which suggest that DL methods struggle to extract high-level seismic features when fed with only a single station record as input, without additional auxiliary information. The lower-than-expected success in our results, aligning with existing literature, underscores the absence of a readily available deep encoder for seismic signals. It is worth noting that early vision encoders required millions of images for training. We believe that a crucial future direction involves gathering an extensive ground or strong motion records dataset (of possibly millions of records) to train DL models, thereby improving accuracy and developing universal seismic deep encoders.

 \clearpage
\clearpage

\section*{Open Research Section}
The raw strong motion waveforms employed in this study, gathered for earthquakes with a magnitude exceeding 3.5, are publicly accessible through the Turkish Disaster and Emergency Management Authority (AFAD) (\url{https://tadas.afad.gov.tr/}). AFAD managed 799 strong motion stations across Turkey, distributed throughout various regions, during dataset compilation. This dataset encompasses strong motion events recorded at these stations since the 1990s, containing detailed information such as occurrence time, epicenter coordinates, depth, magnitude, station number recording the earthquake, station coordinates, and three-channel accelerometer waveforms capturing the entire event in all directions. The dataset utilized in our investigation comprises 36,418 three-channel waveforms corresponding to 3,655 distinct events, gathered from 718 of AFAD's stations over a 8-year period, spanning from January 2, 2012, to December 19, 2018.

\acknowledgments
This work is supported by The Scientific and Technological Research Council of Turkey (TÜBİTAK) as a part of the TÜBİTAK 1001 funded project, Project No.121M732, titled "Deep Learning and Machine Learning Based Dynamic Soil and Earthquake Parameter Estimation Using Strong Ground Motion Station Records”. 

% \bibliography{<name of your .bib file>} don't specify the file extension
% don't specify bibliographystyle

\bibliography{biblo}

\end{document}